\documentclass[fleqn,usenatbib]{mnras}

\usepackage{newtxtext,newtxmath}

\usepackage[T1]{fontenc}

\DeclareRobustCommand{\VAN}[3]{#2}
\let\VANthebibliography\thebibliography
\def\thebibliography{\DeclareRobustCommand{\VAN}[3]{##3}\VANthebibliography}

\usepackage{graphicx}	
\usepackage{amsmath}	

\usepackage{savesym}
\savesymbol{tablewidth}
\usepackage{tabularray}
\restoresymbol{x}{tablewidth}
\usepackage{multirow}
\usepackage[nameinlink]{cleveref}
\usepackage{xcolor}
\usepackage{subcaption}

\newcommand{\eazy}{{\tt{EAZY-py}}}

\newcommand{\pysersic}{{\tt pysersic}}
\newcommand{\bagpipes}{{\tt{Bagpipes}}}

\newcommand{\sextractor}{{\tt{SExtractor}}}

\newcommand{\galfind}{{\tt{GALFIND}}}
\newcommand{\aperpy}{{\tt aperpy}}
\newcommand{\webbpsf}{{\tt{WebbPSF}}}

\newcommand{\cloudy}{{\tt{CLOUDY}}}
\newcommand{\pixedfit}{{\tt{piXedfit}}}
\newcommand{\expanse}{{\tt{EXPANSE}}}
\newcommand{\db}{{\tt{Dense Basis}}}

\def\casgm20{CAS-G-M$_{20}\,$}
\def\m20{M$_{20}\,$}

\title[Behind the Spotlight: Outshining]{Behind the Spotlight: A systematic assessment of outshining using NIRCam medium-bands in the JADES Origins Field}
\author[T. Harvey et al.]{
Thomas Harvey,$^{1}$\thanks{E-mail: thomas.harvey-3@manchester.ac.uk}
Christopher J. Conselice$^{1}$, 
Nathan J. Adams$^{1}$,
Duncan Austin$^{1}$,
Qiong Li$^{1}$,
Vadim Rusakov$^{1}$,\newauthor
Lewi Westcott$^{1}$,
Caio M. Goolsby$^{1}$,
Christopher C. Lovell$^{2}$,
Rachel K. Cochrane$^{3}$,
Aswin P. Vijayan$^{4}$,\newauthor
James Trussler$^{5}$
\\
$^{1}$Jodrell Bank Centre for Astrophysics, University of Manchester, Oxford Road, Manchester M13 9PL, UK\\
$^{2}$Institute of Cosmology and Gravitation, University of Portsmouth
, Burnaby Road, Portsmouth PO1 3FX, UK\\
$^{3}$Institute for Astronomy, University of Edinburgh, Royal Observatory, Blackford Hill, Edinburgh, EH9 3HJ, UK\\
$^{4}$Astronomy Centre, University of Sussex
, Falmer, Brighton BN1 9QH, UK\\
$^{5}$Center for Astrophysics | Harvard \& Smithsonian, 60 Garden Street, Cambridge, MA, 02138, USA
}

\date{Accepted XXX. Received YYY; in original form ZZZ}

\pubyear{2025}

\begin{document}
\label{firstpage}
\pagerange{\pageref{firstpage}--\pageref{lastpage}}
\maketitle

\begin{abstract}
The spatial resolution and sensitivity of JWST's NIRCam instrument has revolutionised our ability to probe the internal structure of early galaxies. By leveraging deep medium-band imaging in the Jades Origins Field, we assemble comprehensive spectral energy distributions (SEDs) using 19 photometric bands for over 200 high-redshift galaxies ($z \geq 4.5$). We present an analysis of this sample with particular emphasis on investigating the "outshining" phenomenon, which can bias the inferred stellar populations by masking the presence of evolved stellar populations ($\geq$ 100 Myr) with the light of bright, young O and B-type stars. We address this problem by performing spatially-resolved SED-fitting of both binned and full pixel-by-pixel photometry, which we compare to the traditional integrated approach. We find evidence for systematic underestimation of stellar mass in low-mass galaxies ($\leq 10^9 \rm M_\odot$) with bursty star formation, which can exceed a factor of 10 in individual cases, but on average is typically a factor of 1.25-2.5, depending on the binning methodology and SFH model used. The observed mass offset correlates with  burstiness (SFR$_{10 \ \rm Myr}$/SFR$_{100 \ \rm Myr}$) and sSFR, such that galaxies with recently rising SFHs have larger mass offsets. The integrated SFH models which produce the most consistent stellar masses are the double power-law and non-parametric `continuity' models, although no integrated model fully reproduces all resolved SFHs. We apply an outshining correction factor to the Stellar Mass Function at $z=7$, finding little impact within the uncertainties. We conclude that outshining can be important in individual low-mass galaxies, but the overall impact is limited and should be considered alongside other systematic SED fitting effects.

\end{abstract}




\section{Introduction}

Obtaining accurate stellar mass estimates at high-redshift has been the subject of many recent studies, since JWST results have suggested there are more galaxies with higher stellar masses than previously predicted \citep{lovell2023extreme, whitler2023star, boylan2023stress, desprez2024lambdacdm}, although there are a lot of other systematic uncertainties around star formation history parametrizations, chosen dust law, stellar population modelling and the initial mass function \citep{lower2020well, Steinhardt2022, Sneppen2022, 2022ApJ...935..146S, woodrum2024jades, wang2024quantifying, jain2024motivation, 2025ApJ...978...89H}. 

The vast majority of photometric studies on high-$z$ galaxy evolution perform unresolved integrated spectral energy distribution (SED) fitting, from fluxes extracted from circular or elliptical apertures. However, with the resolution of JWST's NIRCam ($\sim$0.15\arcsec FWHM in F444W), many galaxies at $z > 5$ are resolved above the PSF \citep[e.g.][]{ormerod2024epochs, 2024arXiv241214970W, varadaraj2024sizes, allen2024galaxy, ward2024evolution, miller2024jwst}, which means a loss of information given the inhomogeneity of high-$z$ galaxies.

One consequence of integrated SED fitting is the underestimation of stellar masses, through a bias known as `outshining' \citep{sawicki1998optical, papovich2001stellar, shapley2001rest, trager2008stellar, graves2010dissecting, maraston2010star,
2018MNRAS.476.1532S, jain2024motivation}. Outshining is thought to occur as a consequence of young, bright O and B-type stars obscuring the presence of older (>100 Myr) stellar populations, which leads to an underestimation of the total mass of stars present in a galaxy. Due to the power-law shape of the IMF, older low-mass stars dominate the total stellar mass but contribute little of the observed rest-UV/optical emission \citep{Schechter1976, Chabrier2000, Kroupa_IMF_2002}.

This can be partially mitigated if the old and young stellar populations do not entirely coincide, but are spatially separated on scales which can be resolved in a binned or pixel-by-pixel analysis of galaxy light. Independent SED fitting of these regions can then recover distinct star formation histories which could otherwise be missed in an integrated analysis. High-redshift, low-mass galaxies are thought to have stochastic star formation histories with episodic bursts of star formation due to shallow potential wells, which may temporarily hide previous generations of stars \citep{pallottini2023stochastic,Looser2023, asada2023bursty, ciesla2024identification,trussler2024epochs, trussler2025like}.

There have been a number of recent applications of resolved SED fitting at high-redshift. \cite{gimenez-artega2022} identified five lensed $5 < z < 9$ galaxies in the SMACS-0723 field and performed pixel-by-pixel SED fitting. They found up to a 1 dex discrepancy between the resolved and integrated stellar mass estimates, which was driven by compact regions of star formation outshining older stellar populations on the outskirts of each galaxy. \cite{2024arXiv240218543F} also observed this effect in the `Cosmic Grapes', a highly magnified ($\mu = 32$) galaxy at $z=6$. As outshining is driven by rapid star formation, the effect is dependent on the specific star formation rate (sSFR) of the galaxy; galaxies with higher sSFRs typically have larger discrepancies between resolved and integrated masses \citep{2018MNRAS.476.1532S, giménezarteaga2024outshining}. \cite{2024arXiv240910963L} studied five $z \sim 5$ bright galaxies with ALMA-CRISTAL observations and did not observe any outshining, with masses consistent within 0.3 dex. They suggested that this is because they probe a higher mass regime (M$_\star = 10^{9.5}$M$_{\odot}$) than studied by \cite{giménezarteaga2024outshining} or \cite{2024arXiv240218543F} ($\leq 10^{8}$M$_\odot$), and these higher-mass galaxies have less stochastic recent star formation histories. \cite{perezgonzalez2022} also finds no stellar mass discrepancy between integrated and resolved methods in $z>3$ red galaxies with M$_\star = 10^{10}$M$_{\odot}$. 

Outshining has also been studied in simulations by \cite{2024ApJ...961...73N}, who propose that a hydrodynamical simulation with a stellar feedback model which leads to episodic `bursty' star formation histories can lead to very significant offsets between the true stellar mass and estimates from SED fitting, including under and over-estimates at low and high masses respectively.
However \cite{cochrane2024highzstellarmassesrecovered} tested stellar mass recovery for galaxies from the SPHINX20 simulations \citep{katz2023sphinx} and instead find reliable recovery of stellar masses using \bagpipes{} within 0.5 dex, even with stochastic star formation histories. They suggest that integrated SED fitting works reasonably well when deep NIRCam wideband observations are available, although they do observe some systematics at low and high stellar masses. \cite{2025arXiv250314591M} have tested spatially resolved recoveries of SFHs for both mock and real observations, finding that traditional integrated SED fitting approaches tend to underestimate the impact of early star formation, and that flexible parametric SFH models perform well to recover stochastic SFHs when applied in a resolved pixel-by-pixel approach.

%

Most early JWST surveys focused on finding high-$z$ galaxy candidates using the most sensitive NIRCam wideband filters, but more recent surveys have expanded this to the numerous medium-band filters, which provide much better constraints on the galaxy SED, allowing more robust redshift and property measurements \citep[e.g.][]{2025arXiv250210282A}. This includes UNCOVER+MEGASCIENCE \citep{2024ApJS..270....7W, suess2024medium}, JADES+JEMS \citep{Eisenstein2023, 2023ApJS..268...64W} and the JADES Origins Field \citep{2023arXiv231012340E,robertson2024earliest}. Medium-band observations are particularly helpful for quantifying the effect of outshining as emission lines and continuum emission can be more readily disentangled, allowing better estimation of past and current star formation activity. 

In this paper we use JWST/NIRCam imaging of the JADES Origins Field, which is one of the deepest fields JWST has observed. When combined with existing HST ACS/WFC imaging, there are 19 photometric bands available, allowing accurate sampling of the galaxy SEDs, which is critical for a resolved analysis. We perform spatially resolved SED fitting with \bagpipes{}, testing multiple pixel binning methodologies in order to determine whether full pixel-by-pixel analysis is necessary. We compare integrated and resolved stellar mass estimates in order to assess the impact of outshining on our galaxy sample. 

This paper is structured as follows. In \autoref{sec:data}
we present the data products, reduction procedure and catalogue creation process. In \autoref{sec:sample_selection} we list our sample selection criteria, and in \autoref{sec:resolved_method} we detail our resolved SED fitting procedure. Our results and discussion, as well as a comparison to literature are presented in \autoref{sec:results}, including detailed analysis of observed outshining in low-mass galaxies and the inferred impact on the galaxy stellar mass function. We conclude and summarise in \autoref{sec:conclusions}.

We assume a standard $\Lambda\mathrm{CDM}$ cosmology with $H_0=70$\,km\,s$^{-1}$\,Mpc$^{-1}$, $\Omega_{\rm M}=0.3$ and $\Omega_{\Lambda} = 0.7$. All magnitudes listed follow the AB magnitude system \citep{Oke1974,Oke1983}.

\section{Data Products and Reduction}
\label{sec:data}

\subsection{JWST/NIRCam Data}

The JADES Origins Field (JOF) is part of the JWST Advanced Deep Extragalactic Survey (JADES), which is a Cycle 1 GTO program. The JOF is within the GOODS South region, and is a parallel field conducted alongside NIRSpec observations of the Hubble Ulta-Deep Field (HUDF). It has been observed with JWST as part of Cycle 1, Cycle 2 and Cycle 3, under program IDs 1180 (P.I. Eisenstein), 1210 (P.I. L{\"u}tzgendorf), 1286 (P.I. L{\"u}tzgendorf), 3215 (P.I. Eisenstein) and 4540 (P.I. Eisenstein), for a total of more than 380 hours with NIRCam, including 15 different medium and wideband filters. 
A full description of the JOF observations can be found in \cite{2023arXiv231012340E}, and an initial search for high-$z$ galaxies is presented in \cite{robertson2024earliest}. 

The NIRCam filters observed are as follows: F090W, F115W, F150W, F162M, F182M, F200W, F210M, F250M, F277W, F300M, F335M, F356W, F410M and F444W. F070W imaging taken in program 4540 was not public when this analysis was conducted. \autoref{fig:depths_areas} illustrates the 5$\sigma$ depths achieved as a function of available area in all filters, demonstrating the depth ($\geq 30$ AB mag) achieved in both medium and wideband filters.

\subsection{HST/ACS WFC Data}

In order to extend the wavelength coverage of our dataset below 0.9~$\mu$m, we incorporate optical ACS/WFC imaging from the \textit{Hubble Space Telescope} (HST).
The JOF has HST ACS/WFC coverage from the Hubble Legacy Fields (HLF) program\footnote{\url{https://archive.stsci.edu/prepds/hlf/}}. We obtain the latest v2.5 data release from MAST for the ACS/WFC imaging in the GOODS-S field \citep{Illingworth2016,Whitaker2019}. Imaging in 5 filters is available; F435W, F606W, F775W, F814W and F850LP, at a $0\farcs03$ pixel scale.

\begin{figure}
    \centering
    \includegraphics[width=\columnwidth]{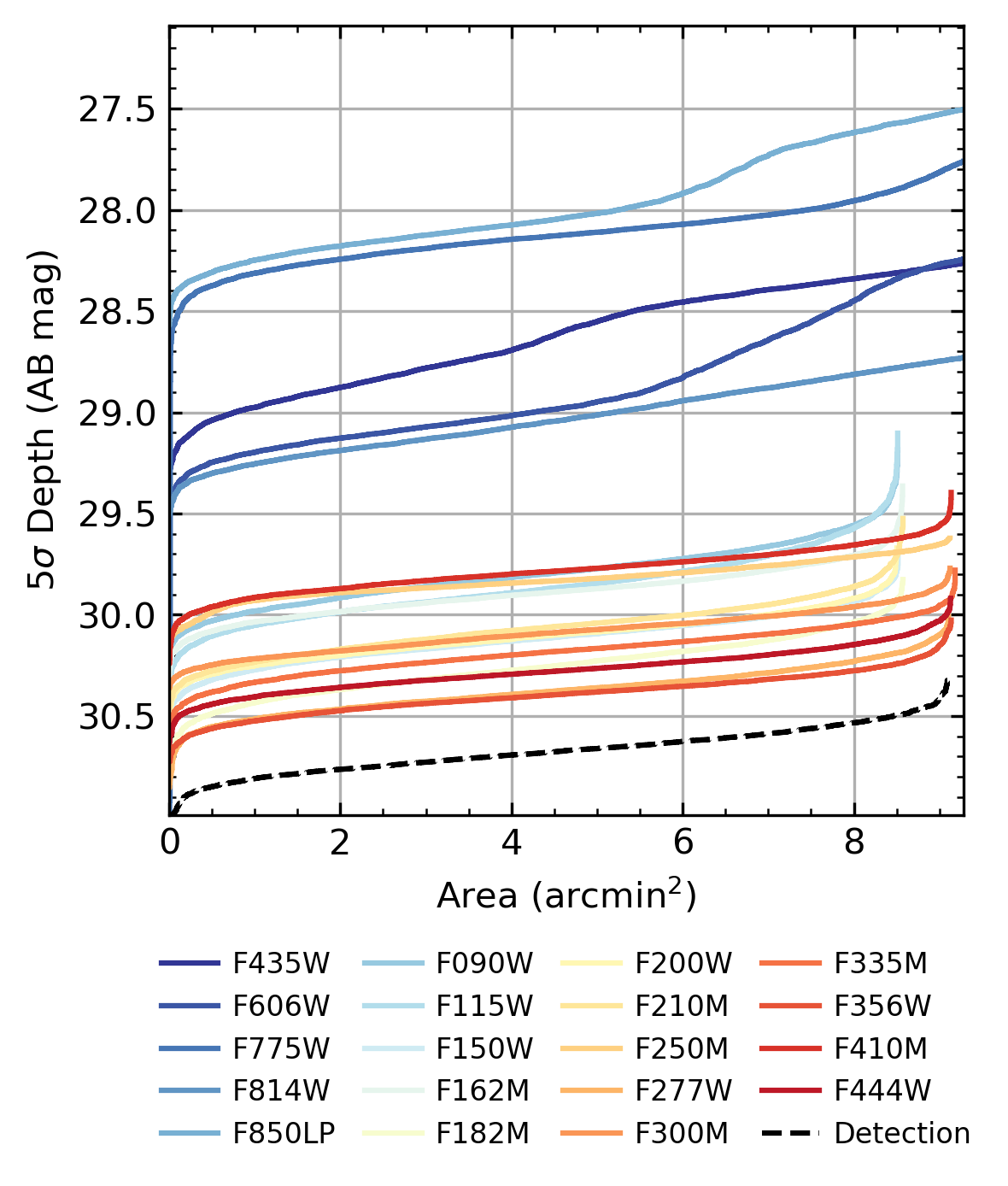}
    \caption{$5\sigma$ depth as a function of observed area in the Jades Origins Field, calculated with $0 \farcs 32$ diameter apertures. The dashed line shows the depth achieved in the detection image, which is an inverse variance weighted stack of the F277W, F356W and F444W images. HST/ACS WFC imaging comes from the Hubble Legacy fields coverage of GOODS-South. Depths in excess of 29.8 AB mags are achieved across the majority of the field in the NIRCam observations. }
    \label{fig:depths_areas}
\end{figure}

We project the ACS/WFC imaging onto the NIRCam imaging World Coordinate System (WCS) using the python package {\tt reproject}\footnote{\url{https://reproject.readthedocs.io/en/stable/}}, using {\tt reproject\_adaptive} with {\tt conserve\_flux = True}. This data is considerably shallower than the NIRCam observations (as seen in \autoref{fig:depths_areas}), but it is useful to constrain Lyman-breaks for bright galaxies at $z<6$.

\subsection{Overview of Reduction Procedure}

Our reduction procedure for the NIRCam imaging follows exactly the procedure described in the EPOCHS paper series \citep{Conselice_2025, adams2024epochs, 2024arXiv240410751A,2025ApJ...978...89H}. In particular \cite{2025arXiv250210282A} details the precise reduction procedure used for the JOF field. 
We use a modified version of the official JWST pipeline, incorporating modifications for improved flat-fielding and artefact removal. We use Calibration Reference Data System (CRDS) pmap1084, along with custom `wisp' templates derived from a medium stack of all surveys reduced within the EPOCHS project (PEARLS, CEERS, JADES, NGDEEP, GLASS) for the F150W, F200W, F182M and F210M filters \citep{2025arXiv250210282A}. We perform a custom 2D background subtraction with {\tt photutils} \citep{larry_bradley_2022_6825092} instead of stage 3's default sky subtraction. Our WCS is derived from Gaia \citep{GAIADR2,GAIADR3} for the F444W image, and then all other filters for both NIRCam and ACS/WFC are matched to the same WCS (using {\tt tweakreg}) and pixel grid using {\tt reproject} with sub-pixel accuracy \citep{robitaille2020reproject}. We note that for the F162M, F182M and F210M filters we chose to re-reduce them with CRDS pmap1210, in order to fix observed image distortion near the corners, as observed in \cite{robertson2024earliest}. We build empirical PSF models for each image by stacking isolated stars, which we detail in Appendix \autoref{sec:psf}.

\subsection{Catalogue Creation}
\label{sec:cat_creation}

Our catalogue creation also follows the EPOCHS paper series methodology, as we use the same \galfind{}\footnote{\url{https://galfind.readthedocs.io/}} package to do our catalogue creation, local depth measurements, initial SED fitting with \eazy{} for redshifts and robust galaxy selection.

Our detection and image segmentation uses \sextractor{}, in two-image mode \citep{Bertin1996}, with fluxes measured in $0\farcs32$ diameter apertures, a minimum area threshold of 9 pixels, a detection threshold of 1.8$\sigma$, and a 2.5 pixel FWHM Gaussian smoothing kernel for detection. All photometry is measured from PSF-matched imaging, as discussed in \autoref{sec:psf}. We derive aperture corrections for each aperture diameter from our F444W PSF model, which we use to correct the measured aperture fluxes. We use the `WHT' extension images for weighting, and disable additional \sextractor{} background subtraction. We also measure total fluxes for each source in Kron apertures, with a Kron factor of 2.5 and minimum radius of 4. For our detection image we use an inverse variance weighted stack of the unhomogenized F277W, F456W and F444W images, in order to optimise the detection of faint galaxies. The depth achieved in this stack is shown by the black dashed line labelled `Detection' in \autoref{fig:depths_areas}. The total unmasked area for the JOF is 8.2 arcmin$^2$, excluding the SW detector gap and large stellar diffraction spikes.

We calculate local depth errors for each galaxy in each image from the Normalised Mean Absolute Deviation (NMAD) of the flux measurement in 200 nearby empty apertures, which we define as $\geq$1$\arcsec$ from any object in the segmentation map. These local depths are more accurate than those produced from \sextractor{} due to correlated noise present in the NIRCam imaging. This aperture flux catalogue is used only for our initial sample selection, and we remeasure total fluxes for selected galaxies as described in \autoref{sec:resolved_method}.



\section{Photo-z Measurements and Sample Selection}
\label{sec:sample_selection}

\subsection{Photo-z Estimation}
\label{sec:photoz}

We perform initial photo-$z$ estimation with \eazy{}, following the same methodology as the EPOCHS papers \citep{adams2024epochs, Conselice_2025}. We use \eazy{} \citep{brammer2008eazy} with the default templates (tweak\_fsps\_QSF\_12\_v3), along with Set 1 and Set 4 of the SED templates generated by \cite{Larson2022}. These additional templates were developed to extend the colour space covered by the default templates to include bluer rest-frame UV colours as well as stronger emission lines, both of which have been observed in high-redshift galaxies. Our initial photo-$z$ estimation is done using aperture corrected fluxes with a minimum flux uncertainty floor of 10\% to account for potential flux calibration and zero-point issues as well as intrinsic template inaccuracies, and we do not allow \eazy{} to iteratively tune zeropoints, which could vary between detectors, or to apply its default priors on the UV $\beta$-slope or apparent magnitude. We allow the redshift to vary with $0 \leq z_{\rm phot} \leq 25$ in the primary run, and also perform multiple other runs with lower photo-$z$ limits, in order to have a low-$z$ comparison available for all galaxies in our sample. This includes runs with $z_{\rm max} = 2, 4, 6$. For each robust candidate we perform a $\chi^2$ test between the primary run and the low-$z$ run where $z - z_{\rm max} > 0.5$ as part of our selection criteria. 

\subsection{Selection Criteria}

Our selection criteria for robust high-$z$ galaxy candidates follows that of the EPOCHS paper series \cite{adams2024epochs,Conselice_2025,2024arXiv240410751A}. 

In short, we require the following:
\begin{enumerate}
    \item Spurious Source Criteria: 
        \begin{enumerate}
            \item Unmasked in all photometric bands.
            \item 50\% enclosed flux $>$ 1.5 pixels in LW NIRCam photometry, to avoid spuriously selecting hot pixels.
        \end{enumerate}
    \item SNR requirements: 
        \begin{enumerate}
            \item $<3\sigma$ detection in all bands below the 1216\AA{} Lyman break, with at least one band covering this region of the SED, given the \eazy{} photo-$z$ estimate. 
            \item $>5\sigma$ detections in the first two wideband NIRCam filters with filter transmissions fully redward of the Lyman break and $>2\sigma$ detections in the other remaining redward wideband NIRCam filters. 
        \end{enumerate}
    \item Photo-$z$ Reliability Requirements: 
        \begin{enumerate}
            \item Maximum likelihood EAZY redshift $z > 4.5$. 
            \item The primary photo-$z$ PDF integral must satisfy $\int^{1.10\times z_{\textrm phot}}_{0.90\times z_{\textrm phot}} \ P(z) \ dz \ \geq \ 0.6 $ to ensure that the redshift is strongly constrained. $z_{\textrm phot}$ refers to the redshift with maximum likelihood from the \eazy{} redshift posterior. 
            \item We require the best-fitting \eazy{} SED to satisfy $\chi^2_{\rm red} < 3 (6)$ to be classed as a robust (good) fit.
            \item We require a difference of $\Delta \chi^2 \geq$ 4 between the high-$z$ and selected low-$z$ \eazy{} run for each galaxy, ensuring the high-$z$ solution provides a significant statistical improvement to the fit. 
        \end{enumerate}
    \item Brown Dwarf Removal:
        \begin{enumerate}
            \item For compact sources (\sextractor{} {\tt FLUX\_RADIUS}$<$ F444W PSF FWHM) we then require that $\Delta \chi^2 \geq$ 4 between the best-fitting high-$z$ galaxy solution and the best-fitting brown dwarf template. All compact sources are fitted with photometry derived from all available Sonora Bobcat and Cholla brown dwarf models \citep{marley_mark_2021_5063476, karalidi2021sonora}, using the fitted template with lowest $\chi^2$ as the best model for each galaxy.
        \end{enumerate}
\end{enumerate}

Our brown dwarf fitting code is available on GitHub as \textit{BD-Finder}\footnote{\url{https://github.com/tHarvey303/BD-Finder/}} and will fit Sonora Bobcat, Cholla, Diamondback and Elf Owl brown dwarf models \citep{marley_mark_2021_5063476, karalidi2021sonora, morley2024sonora, mukherjee2024sonora} as well as the low-metallicity LOW-Z models of \cite{meisner2021new}. Our JOF brown dwarf candidates will be released in an upcoming paper. 

\begin{figure}
    \centering
    \includegraphics[width=\columnwidth]{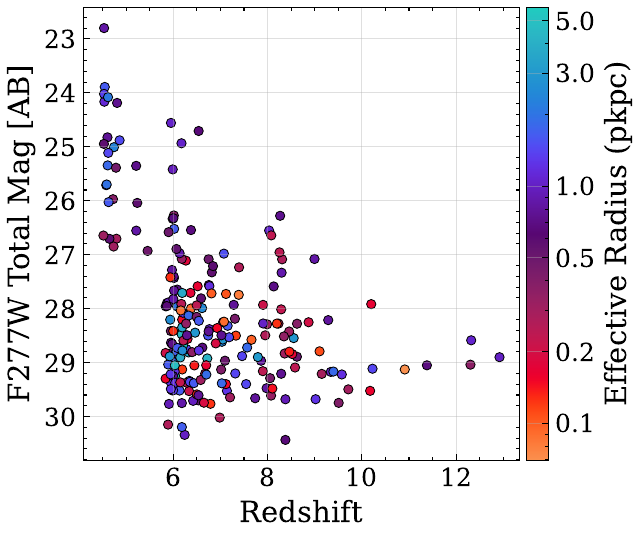}
    \caption{Photometric redshift from \eazy{} vs observed F277W total magnitude for our high-z galaxy sample, coloured by the effective radius as measured by single Sérsic fitting using \pysersic{}. The lack of faint galaxies at $z \leq 6$ is a product of the shallower depth of the HST observations we use in our selection procedure, and we include bright galaxies at $z\leq6$ only to test our resolved analysis technique on bright, massive galaxies. 
    The total magnitude is calculated by correcting the $0\farcs32$ aperture flux for both the total extent of the galaxy as measured by the \sextractor{} Kron aperture as well as the PSF correction derived from our F444W PSF model given the Kron aperture. The Kron radius is calculated from the \sextractor{} output parameters as $\tt{KRON\_RADIUS}\times \sqrt{\tt{A\_IMAGE}\times B\_IMAGE}$.}
    \label{fig:kron_vs_radius}
\end{figure}

This selection criteria results in 222 galaxy candidates at $z>4.5$ selected from 16,335 objects in the full catalogue. We set our lower redshift limit to $z=4.5$ to ensure we can constrain the Lyman-break for all selected galaxies. Due to our requirement of a 5$\sigma$ detection in the first two bands redward of the break, our sample completeness at $4.5 < z < 6$, where these bands fall in the shallow ACS/WFC data, is significantly poorer than at $z>6$, as can be inferred from \autoref{fig:kron_vs_radius}. \cite{2025arXiv250210282A} provides a discussion of the accuracy of our selection criteria and photo-$z$ estimation in the JOF field with a comparison to available spectroscopic redshifts. Our sample differs slightly from that used in \cite{2025arXiv250210282A}, due to our inclusion of the HST ACS/WFC data during our SED fitting and selection procedure. 

\subsection{Galaxy Sizes}

We use \pysersic{} to measure galaxy sizes, by fitting a single Sérsic profile to each galaxy in the F444W band. \pysersic{} is a Bayesian tool for measuring structural galaxy parameters and morphologies \citep{pasha2023pysersic}. We mask other sources within the cutouts during the fitting using the segmentation map. We use the gradient MCMC No U-turn Sampler \citep[NUTS, ][]{hoffman2011no} via {\tt numpyro} \citep{phan2019composable}, to robustly sample the posterior and constrain the Sérsic parameters. We place a prior on the position of the Sérsic profile to constrain it to within 5 pixels of the center of the cutouts. We mask other sources in the fitting, as defined from the segmentation map.

An overview of our sample is shown in \autoref{fig:kron_vs_radius}, where we display the total F277W magnitude (as described in \autoref{sec:cat_creation}) vs the EAZY photometric redshift for each galaxy, coloured by the effective radius in kpc for each galaxy from the \pysersic{} fitting.

\section{Resolved SED-Fitting Methodology}
\label{sec:resolved_method}

In this section we describe our process for resolved SED fitting. This is performed through our Python package \textit{EXtended Pixel-resolved ANalysis of SEDs} (\expanse{}), which we release publicly on GitHub \footnote{\url{https://github.com/tHarvey303/EXPANSE}}. \expanse{} performs PSF modelling and convolution, cutout creation, pixel binning, SED fitting and includes an interactive web-based app for the analysis of results. It is designed to be flexible, and works with multiple pixel binning methods and multiple SED fitting tools, including \bagpipes{}, and \db{}. It also ties in with its sister package \galfind{}\footnote{\url{https://galfind.readthedocs.io}} for easy configuration of image paths, catalogue information and cutout creation, but can also be used entirely independently given a set of input images, as well as a galaxy position and cutout size. An overview of \expanse{} is given in Appendix \autoref{sec:expanse}.

We produce square cutouts of our galaxy sample in all bands, centered on the \sextractor{} coordinates. The cutout size is set to be 2.5$\times$ the Kron radius, with a minimum size of $65 \times 65$ pixels ($1.96\arcsec \times 1.96\arcsec$). We define the contiguous region occupied by each galaxy from the segmentation map in the detection image. In the case of nearby neighbours, or where the segmentation map does not appear to accurately describe the extent of the galaxy, we compute a new segmentation maps using the Python package ${\tt sep}$ \cite{Bertin1996, barbary2016sep}. 
We note that the borders of the segmentation maps can sometimes be non-contiguous, but we make no attempt to smooth these features as the fraction of the total flux enclosed in these outlying regions is typically very small. 

\subsection{Pixel Binning}
\label{sec:binning}
Whilst it is possible to perform full pixel-by-pixel SED fitting for a galaxy, this approach requires significant computational resources given the high resolution of images produced by NIRCam. Given that for NIRCam the PSF size in all bands is significantly larger than the pixel scale ($0\farcs03$), it is also potentially misleading to infer differences in SED or derived parameters on the scale of individual pixel elements. The reliability of results derived from pixels with low signal to noise can also be uncertain. 

In this work we instead opt to test a number of different pixel binning approaches, as well as a full pixel-by-pixel analysis, in order to understand the systematic differences introduced by the choice of binning procedure. 

Firstly we employ the pixel-binning methodology of \pixedfit{} \cite[]{2017MNRAS.469.2806A,2021ApJS..254...15A,2022ApJ...926...81A,2022ApJ...935...98A,2023ApJ...945..117A}. In brief, this binning approach aims to segment a galaxy into multiple components whilst ensuring each component reaches a minimum SNR and diameter to ensure reliable photometry. An iterative approach is taken, with the first bin centered on the brightest pixel in a reference band. The SEDs of neighbouring pixels are compared using a $\chi^2$ calculation such that only neighbouring pixels which meet a set $\chi^2$ threshold can be combined. Pixels which meet these threshold are added to the bin until the bin meets the SNR and diameter requirements. The process then repeats from the next brightest pixel which has not yet been binned, and continues until all pixels within the galaxy region are assigned to a bin. This approach is designed to maximize the spatial SED information provided by the high resolution NIRCam images, whilst also respecting the limits imposed by the PSF and ensuring a minimum signal to noise within each bin.

Our first binning method requires a minimum SNR of 7 in all bins across all the NIRCam filters which cover wavelengths beyond the Lyman break. The observed wavelength of the Lyman break is derived from the \eazy{} SED-fitting to the aperture photometry for each galaxy, as discussed in \autoref{sec:photoz}. We do not impose a minimum SNR requirement in any of the HST bands, as they have much lower depths that the NIRCam imaging, as can be seen in \autoref{fig:depths_areas}. We will refer to this as \textbf{`pixedfit'} binning.

We use the F277W band as the reference band to place the initial bin on the brightest pixel within the galaxy, as it is typically one of the deepest observation for each galaxy.  We set a $\chi^2_r \leq 5$ limit between pixels when comparing  photometry in all bands. The minimum extent of a single bin is set to 7 pixels (0\farcs21), in order to avoid segmenting significantly below the limits imposed by the NIRCam PSF. 

Our other binning procedures are designed to provide significantly larger numbers of bins for a given galaxy. We do not apply them to every galaxy in the sample, but instead limit them to the low-mass regime (< 10$^9$ M$_\odot$) as measured from our \bagpipes{} SED fitting discussed in \autoref{sec:sed_fitting}. As discussed later in \autoref{sec:outshining}, this is because we see consistent stellar masses between our resolved and integrated fitting with our initial \pixedfit{} binning in the mass range.

Our second binning procedure is also based on the \pixedfit{} algorithm, but with a minimum diameter of 2 pixels and a lower SNR requirement per bin of 4, which we will refer to as `\textbf{pixedfit nomin'} binning.

The third test we conduct is replacing the \pixedfit{} binning algorithm with Voronoi tessellation, as implemented in {\tt{vorbin}} by \cite{Cappellari2003}. We require a minimum SNR/bin in the F277W band of 7, to approximately match our SNR requirement for \pixedfit{} binning. We find a typical bin-to-bin scatter in achieved SNR of 10-25\%. As Voronoi tessellation will not group individual pixels which meet SNR thresholds, the centres of objects are typically fit in a direct pixel-by-pixel manner, resulting in higher numbers of bins per galaxy. We will refer to this as \textbf{`Voronoi'} binning.

Our final binning procedure is a full pixel-by-pixel analysis, where every pixel meeting a set SNR criteria is fit individually. We apply a criteria similar to \cite{giménezarteaga2024outshining}, requiring a 2$\sigma$ per pixel detection in all NIRCam widebands redward of the Lyman break to include a pixel in the binned region. We note that this requirement means that some pixels within our full galaxy region may not be fit at all, as they do not reach the SNR criteria. We correct for this by scaling our derived parameters (stellar mass, SFR) by the ratio of the sum of the light within the binned pixels to the total galaxy light within the entire galaxy region, including the low SNR pixels, using the F444W image. This allow for a fair comparison of these parameters with our integrated SED fitting results. We will refer to this as \textbf{pixel-by-pixel} binning. 

\begin{figure}
        \centering
        \includegraphics[width=\columnwidth]{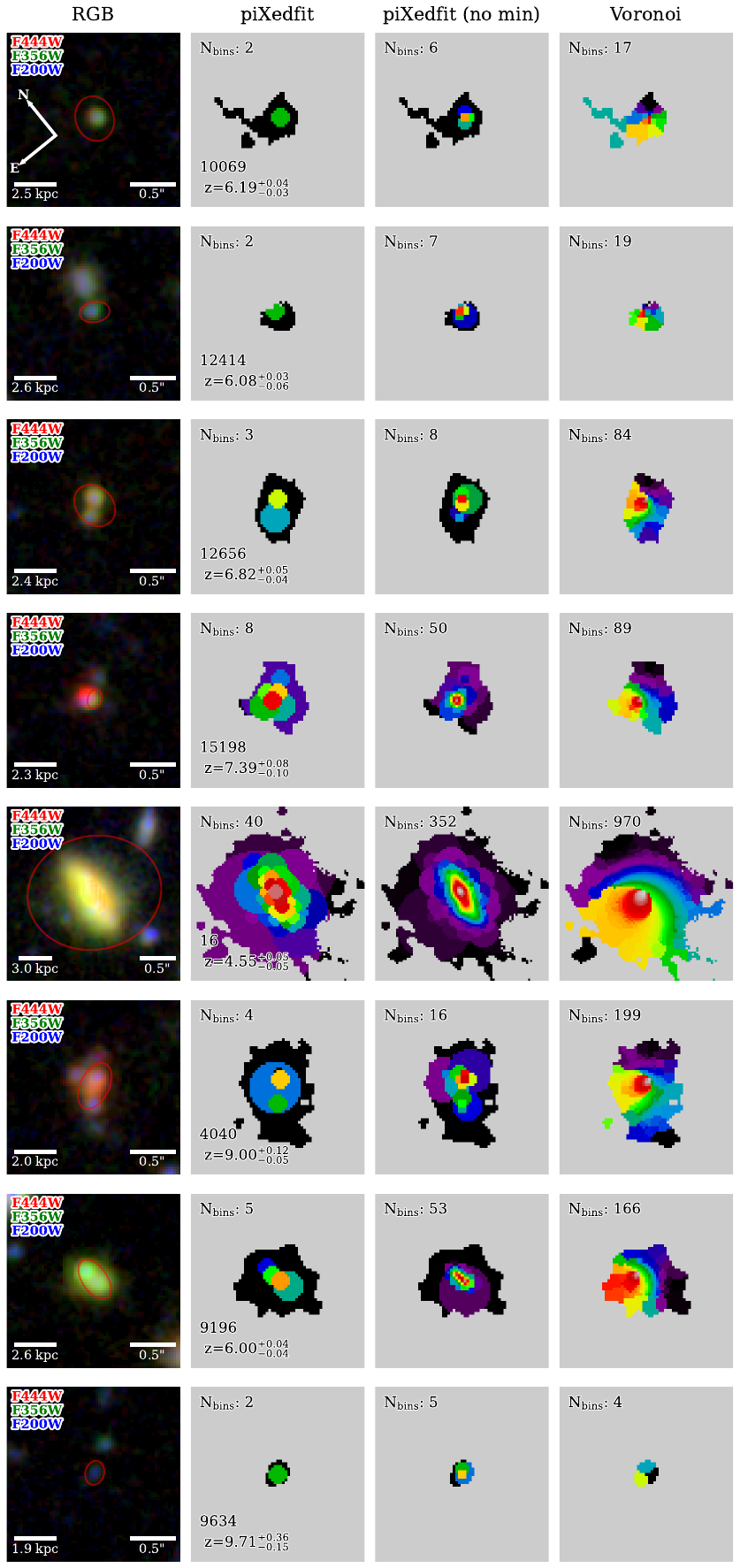}
        \caption{Examples of binned regions for a subset of our galaxy sample at $4.5 \leq z \leq 10.5$ for three different binning methodologies, alongside an RGB image of each galaxy scaled using the \protect\cite{lupton2004preparing} method. Each bin is shown in a different colour for clarity. }
        \label{fig:binning_methods}

\end{figure}

\autoref{fig:binning_methods} shows RGB cutouts (scaled by the method of \citealt{lupton2004preparing}) of randomly chosen galaxies from the sample at a range of sizes and redshifts, along with the extent of the galaxy region and the binmaps for each binning technique, where each bin is shown in a different colour, and the background is shown in grey. In all cases the number of bins increase from left to right, typically by a factor of 5 to 10. 

If a given binning procedure on a given galaxy does not produce two or more bins for a galaxy, we do not include that galaxy in further analysis using that binning method. For the initial \pixedfit{} binning, 133 galaxies are binned into two or more bins. For the `\pixedfit{} nomin' binning, 176 galaxies are binned into two or more bins and for the Voronoi binning, 195 galaxies are binned into two or more bins. 

\autoref{tab:bin_stats} gives the statistics for each binning method, including the mean and median number of bins each galaxy is split into, as well as the maximum and the total number of bins for all galaxies. We also give the number of galaxies for each method which are not broken into multiple bins and hence excluded from the analysis. Galaxies which do not meet the binning criteria are typically compact sources close to the overall SNR limit. 

\begin{table}
\caption{Table showing statistics for each binning method tested, including the median, mean, and maximum number of bins fitted to each individual galaxy using each method. We also list the total number of fits performed for each binning method as well as the number of galaxies where the galaxy was split into either one or no bins, and hence excluded from the resultant analysis for that binning method.}

\begin{tabular}{l|ccccc}
\textbf{Binning Method} & \textbf{Median} & \textbf{Mean} & \textbf{Max} & \textbf{Total} & \textbf{Num N}$_{\rm bin} \leq 1$ \\ \hline
`pixedfit' & 2.0 & 6.2 & 79 & 829 & 89 \\
`pixedfit nomin' & 6.0 & 37.7 & 889 & 6636 & 46 \\
`Voronoi' & 16.0 & 74.1 & 1413 & 14750 & 23 \\
`pixel by pixel' & 32.0 & 74.3 & 1132 & 12118 & 59 \\
\end{tabular}
\label{tab:bin_stats}
\end{table}

\subsection{Binned and Integrated Photometry}
We measure the photometry in each bin from each PSF-homogenized cutout for every filter for each galaxy, with uncertainties derived from the RMS error maps produced as part of the reduction procedure, which we add in quadrature. On a subset of our galaxy images we test whether the RMS error maps capture the true noise present within the images, by calculating the per-pixel RMS of blank regions of the cutouts. We find that the RMS maps typically underestimate the flux uncertainties in each band by 20-40\%, which we correct for in our uncertainty estimates by applying a per/band correction factor to the RMS of each pixel when measuring photometry. We note that during SED fitting we also apply an additional error floor of 5\% to account for any residual uncertainties in the flux or limitations within the modelling itself. 
The flux used for our integrated SED fitting for comparison is simply the sum of all pixels within the galaxy region, with appropriate error propagation. Note that we do not perform any aperture correction to the integrated or resolved photometry, or correct for any flux not captured within the galaxy region. As we are interested in comparing the like-to-like results of the SED fitting, where the total flux is preserved, and as our photometry is measured from PSF-matched imaging, this effects only the overall normalization of the SED fitting, rather than changing the relative photometry between filters. 

\subsection{Resolved SED Fitting}
\label{sec:sed_fitting}

We perform SED fitting for both the integrated and resolved photometry using a modified version of \bagpipes{} v1.2.0, in a manner similar to that of \cite{2025ApJ...978...89H}.  Our modifications of Bagpipes do not affect the fitting procedure or models, but simply provide more functionality and additional derived parameters such as emission line equivalent widths, fluxes and UV measurements such as $\beta$ slope and M$_{\rm UV}$.

The parameters, priors and hyper-parameters of our chosen models are listed in \autoref{tab:bagpipes_table}. We employ \textit{Binary Population and Spectral Synthesis} \citep[BPASS,][]{stanway2018re} SPS models rather than the default \cite{bruzual_charlot_2003} models, which account for the presence of binary stellar populations. Specifically we use models generated with v2.2.1 of BPASS, with the default IMF (power-law slope of 1.35, upper mass cutoff of 300 M$_{\odot}$). 

We include emission lines and nebular continuum based on \cloudy{} v17.03 \citep{2017RMxAA..53..385F}. We regenerate \cloudy{} models in order to probe a wider range of the ionisation parameter U between -3 $\leq \log_{10} U \leq $ -1 using a \cloudy{} configuration file distributed with \bagpipes.

We use the \cite{calzetti2000dust} prescription for dust as it was found by \cite{2024MNRAS.527.5808B} that UV-selected high-$z$ galaxies in the ALMA REBELS survey follow the local Calzetti-like IRX-$\beta$ relation, so we do not fit a more complex dust law. The allowed stellar metallicity ranges from  $10^{-3} \leq Z_\star/Z_\odot \leq 2.5$ with a uniform prior, as these galaxies are expected  to have low metallicity, but theoretically could enrich their local environments quickly \citep{langeroodi2023evolution,curti2023chemical}. 

For our integrated photometry fits, we fix the redshift prior to the PDF from our \eazy{} SED-fitting, which we approximate as a Gaussian. The redshift prior draws are capped at $\pm 3 \sigma$. For the resolved photometry fits we do not allow the redshift to vary, due to the lower SNR of the individual bins, so all bins for a galaxy are fixed to the best-fit redshift derived from the integrated fit using the delayed-exponential SFH. We also test fixing our integrated photometry fits to the same fixed redshift as the resolved fits and find it has no overall effect on the results.  We use the default sampling method, using the Python package {\tt PyMultiNest} \citep{feroz2009multinest, buchner2016pymultinest}. 

The SFH models listed in \autoref{tab:bagpipes_table} are the models used for our integrated SED fitting. We test 7 different SFH models, including 4 parametric models and 3 non-parametric models. The parametric models are the log-normal, delayed-exponential, constant, and double power-law SFH models. The constant model is a simplistic flat SFR model with a fitted time for SF to begin and end. The other three are smoothly varying SFH models which peak at a given lookback time. Their functional forms are described in \cite{carnall2019measure}, but all the models tested allow for periods of rising and declining SFR. The non-parametric models include two models with fixed time bins, where the ratio of SFR between bins is fitted, which are first described in \cite{leja2019measure} and \cite{tachella2022}. We fit two variations of this `continuity' model with a different choice of hyper-parameter for the Student's-t prior describing the SFR bin ratios. The second model we call the `continuity bursty' SFH model because larger SFR ratios are allowed, which allow for more stochastic star formation histories. We also fit the dense basis model of \cite{Iyer2019NonparametricFormation}, which uses gaussian processes to construct smoothly varying star formation histories without time bins, by constraining the fraction of total stellar-mass formed in different fractions of available time for star formation. 
We have not attempted to collate all SFH models in use, but we have aimed to test commonly used models in the literature in order to test which perform best when compared to our resolved fits. 

For the resolved SED fitting (i.e. the fits to each individual bin/pixel) we test two models: the constant star formation history model and the non-parametric `continuity bursty' model described previously. Both of these are configured as described in \autoref{tab:bagpipes_table}. A constant SFH is quite simplistic, but given that each galaxy is fit by multiple bins, the overall composite SFH can be significantly more flexible. The continuity bursty model is also tested to examine the variation in stellar mass caused by the assumed SFH for each individual bin, where a more complex SFH may be required even for a small region within a galaxy.

\begin{table*}
\centering
\caption{Summary of parameters, hyper-parameters and priors for our \bagpipes{} SED-fitting. Parameters and priors for other iterations can be assumed to be the same as given for the `fiducial' bagpipes run unless otherwise specified. The top section of the table lists parameters that are common to all of our \bagpipes{} models, whereas the lower section gives the model-specific parameters for each of our chosen configurations. \label{tab:bagpipes_table}}
\begin{tblr}{
cell{1}{1} = {c=2}{},
}

\textbf{Common Parameters} & &  \\ \hline \hline
Parameter                             & Prior/Value (Min, Max)                               & Description                         \\ \hline
SPS Model                             & BPASS v2.2.1                                   & Stellar population synthesis model  \\
IMF                                   & default BPASS IMF               & Stellar Initial Mass Function       \\
Dust Law parametrization                            & \cite{calzetti2000dust} & Dust law                            \\
A$_\textrm{V}$ & uniform: (0, 5);  & V-band attenuation (all stars) \\

$\log_{10}(M_\star/M_{\odot})$            & uniform: (5, 12)                   & Surviving stellar mass              \\
Z$_\star/\textrm{Z}_{\odot}$              & uniform: (0.001, 2.5);                                    & stellar metallicity                 \\
Z$_{\textrm{gas}}/\textrm{Z}_{\odot}$ &  Fixed to Z$_\star$                          & gas-phase metallicity               \\
$\log_{10}\textrm{U}$                 & uniform: (-3, -1)               & Ionization Parameter                \\
\hline
\end{tblr}

\begin{tblr}{cell{1}{1} = {c=2}{}}
\textbf{Integrated SFH Parameters} & & \\  \hline \hline

Model & Parameter                             & Prior/Value (Min, Max)                               & Description                         \\ \hline
All & $z_{\textrm{phot}}$                   & \eazy{} Posterior PDF ($\pm 3 \sigma$)                      & Redshift                            \\
Log-normal SFH & {t$_\textrm{max}$} & uniform: (10 Myr, 15 Gyr) & Age of Universe at peak SFR \\
& FWHM & uniform: (10 Myr, 15 Gyr)  & FWHM of SFH \\
delayed-$\tau$ SFH & $\tau$ &        uniform: (10 Myr, 15 Gyr) & e-folding timescale  \\
& Age & log-uniform:  (10 Myr, t$_{\rm univ}(z_{\rm phot}))$ & Time since SF began \\
constant SFH & Age min & uniform (0, 2.5 Gyr) & Time SFH started \\
&  Age max & uniform (0.01 Gyr, 2.5 Gyr) & Time SFH ended \\
double power-law SFH & $\tau$ & uniform (0.001 Gyr, 3 Gyr) & Time of Peak Star Formation\\
& $\alpha$ & uniform (0, 100) & Falling power-law slope \\
& $\beta$ & uniform (0, 100) & Rising power-law slope \\
`continuity' & N$_{\textrm{bins}}$ & 6 bins (5 fitted parameters) & First bin 0 -- 10 Myr, SF begins at $z = 20$,  \\
non-parametric SFH & & & others distributed equally in log$_{10}$ lookback time \\
& d$_{\log_{10}\textrm{SFR}}$ & Student's-t: $\nu = 2, \sigma = 0.3 $ & Ratio of $\log_{10}$SFR in adjacent bins, coupled by $\sigma$ \\
`continuity bursty'  & N$_{\textrm{bins}}$ & 6 bins (5 fitted parameters) & First bin 0 -- 10 Myr, SF begins at $z = 20$,  \\
non-parametric SFH & & & others distributed equally in log$_{10}$ lookback time \\
& d$_{\log_{10}\textrm{SFR}}$ & Student's-t: $\nu = 2, \sigma = 1.0 $ & Ratio of $\log_{10}$SFR in adjacent bins, coupled by $\sigma$ \\
\cite{Iyer2019NonparametricFormation} & N$_{\rm bins}$  & 4 & Number of fitted SFH percentiles  \\
& $\alpha$ & 3 & Dirichlet parameter for SFH \\
\hline
\end{tblr}
\end{table*}

\subsection{Rest UV and Emission Line Properties from Photometry}
\label{sec:emission_line_properties}
We follow the methodology of \cite{2024arXiv240410751A} to estimate the UV properties of our galaxy sample in a spatially resolved manner. With the same bins used in the SED fitting we measure the UV slope $\beta$, where $f_{\lambda} \propto \lambda^{\beta}$ by directly fitting a power-law to the photometry, which we shift to the rest-frame using the \eazy{} derived photo-$z$. We limit the fitted photometry to filter bandpasses enclosed within the range 1250, 3000\AA{} to avoid bias from rest-UV emission lines. 

We also measure m$_\textrm{UV}$ and M$_\textrm{UV}$ in a tophat between 1450\AA{} and 1550\AA{} restframe. We do not attempt to correct individual bins for the biases discussed in \cite{2024arXiv240410751A}, which can cause systematic offsets of $\Delta\beta\simeq -0.55$ at low SNRs.

Given the wide and medium band filter observation available, at specific redshift ranges we use these observations to estimate emission line equivalent widths by inferring the relative strengths of the line and continuum from a photometric excess in the filters such that:

\begin{equation}
\Delta \mathrm{m}=-2.5 \log_{10} \left(1+\frac{E W_{\text {Sum }}(1+z)}{\text { Bandwidth }}\right)
\end{equation}
where $E W_{\rm Sum}$ is the equivalent width of all emission lines in the filter, Bandwidth is the width of the filter, and $\Delta \mathrm{m}$ is the measured photometric strength. This method is only reliable for strong lines which dominate nearby emission features, such as H$\alpha$. We estimate H$\alpha$ equivalent widths for galaxies at $z \leq 6.6$, as at higher redshifts H$\alpha$ emission is not observable with our NIRCam filterset. We estimate full equivalent width posteriors from which we estimate uncertainties through a Monte Carlo simulation given the photometric uncertainties on the relevant filters used. 

\cite{duan2024adding} demonstrate that this method can recover broadly accurate equivalent widths but is limited by systematic underestimation of $30\% \pm 20\%$ due to overestimation of the continuum level. We are interested only in the broad correlation of the EW with other properties, rather than the accuracy of individual measurements, so this technique is sufficient for this purpose. 

\section{Results and Discussion}
\label{sec:results}

In this section we present the results of our analyses of the spatially resolved properties of our galaxy sample, and discuss the implications of these results with a comparison to the literature. We focus primarily in this work on the application of our resolved SED fitting procedure to better constrain galaxy stellar mass estimates, and leave exploration of other resolved properties to a future work, although we do briefly discuss resolved vs integrated star formation rates in Appendix \autoref{sec:resolved_sfr}.

\subsection{Property Maps from Resolved SED Fitting}

We performed SED fitting on both the integrated and resolved SED fitting as described in \autoref{sec:sed_fitting}. This allows us to build maps of derived properties, including stellar mass and star formation rate (or corresponding surface densities). For stellar mass and SFR specifically we take a further step of `supersampling' the derived maps by assuming that the mass and SFR are directly proportional to the flux in F444W and the band closest to 1500\AA{} rest-frame, respectively. This lets us present these maps weighted by the flux in these filters, allowing recovery of finer detail in the mass and star formation maps. 

\begin{figure*}
    \centering
    \includegraphics[width=\textwidth]{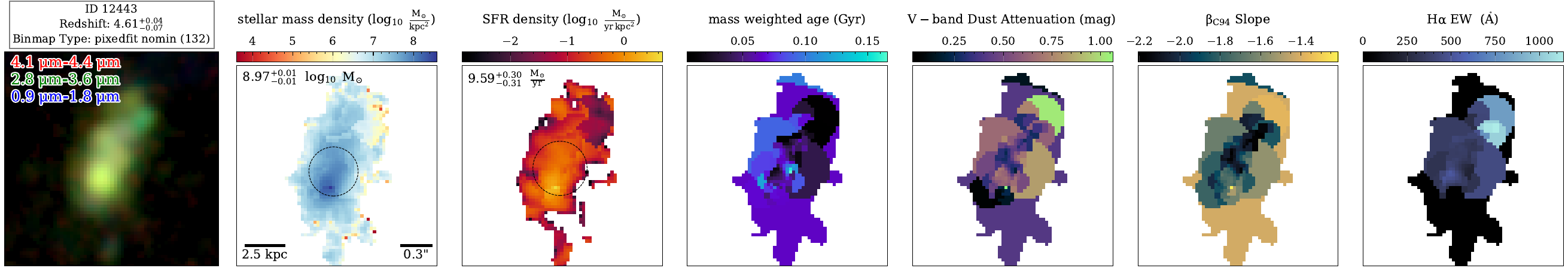}
     \includegraphics[width=\textwidth]{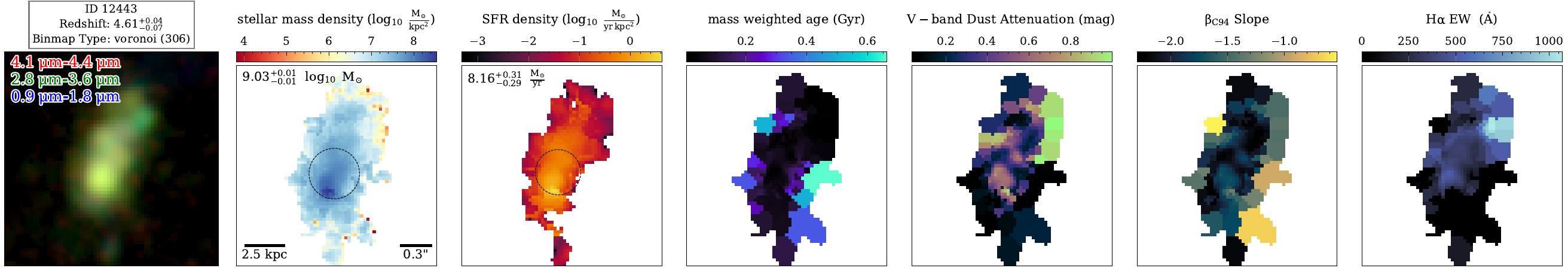}
    \includegraphics[width=\textwidth]{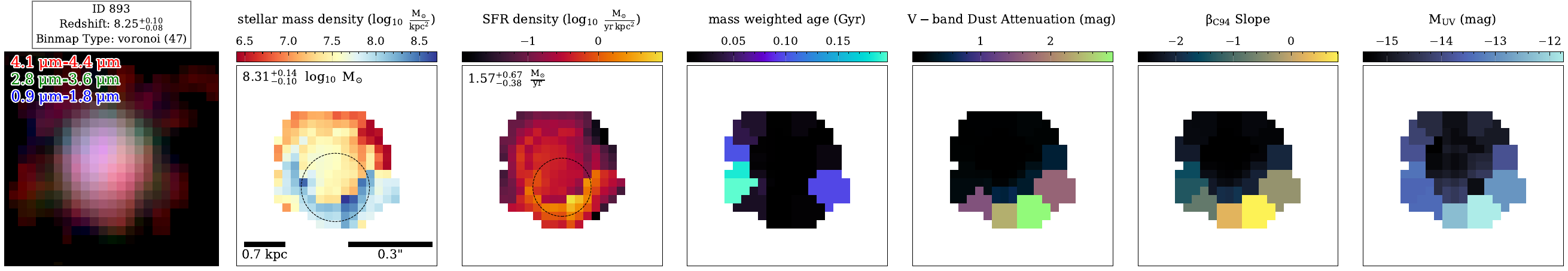}
    \includegraphics[width=\textwidth]{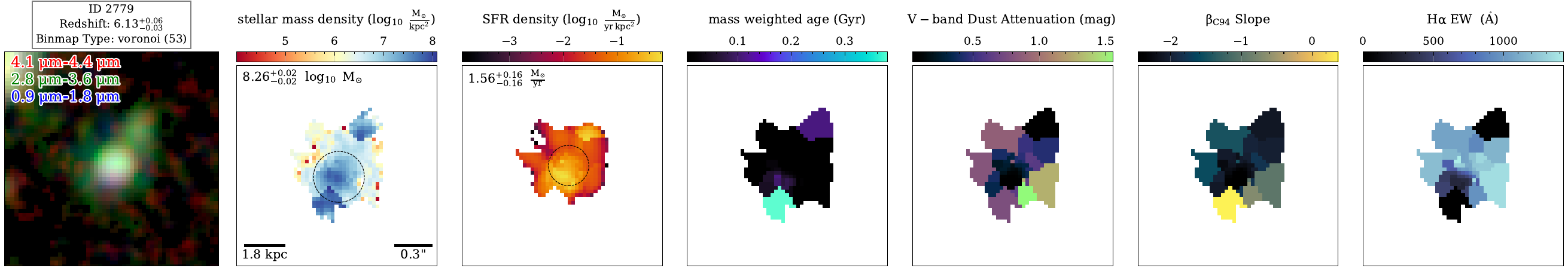}
    \includegraphics[width=\textwidth]{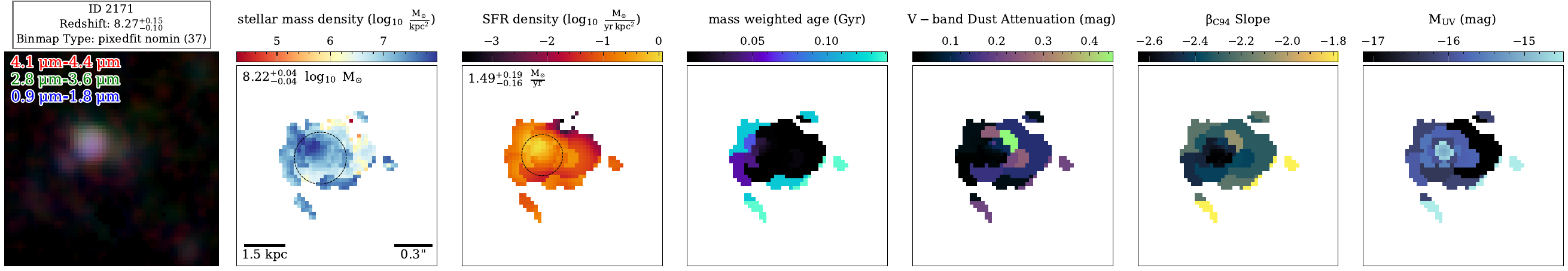}
    \includegraphics[width=\textwidth]{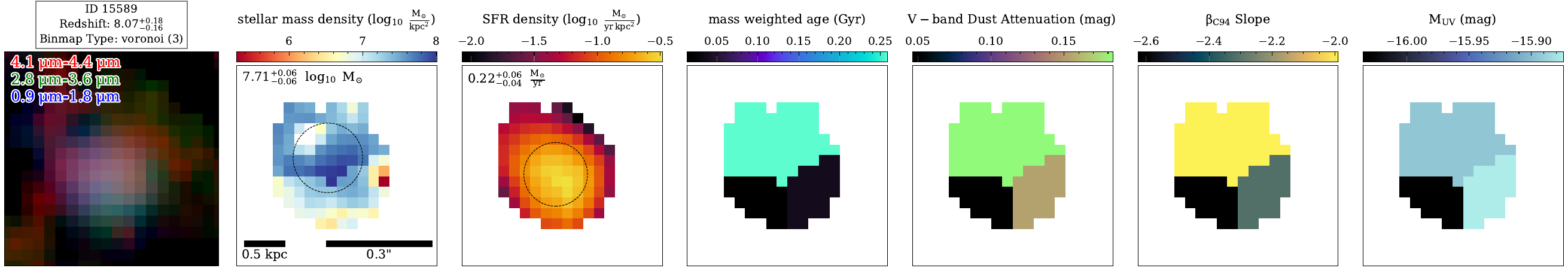}
    \includegraphics[width=\textwidth]{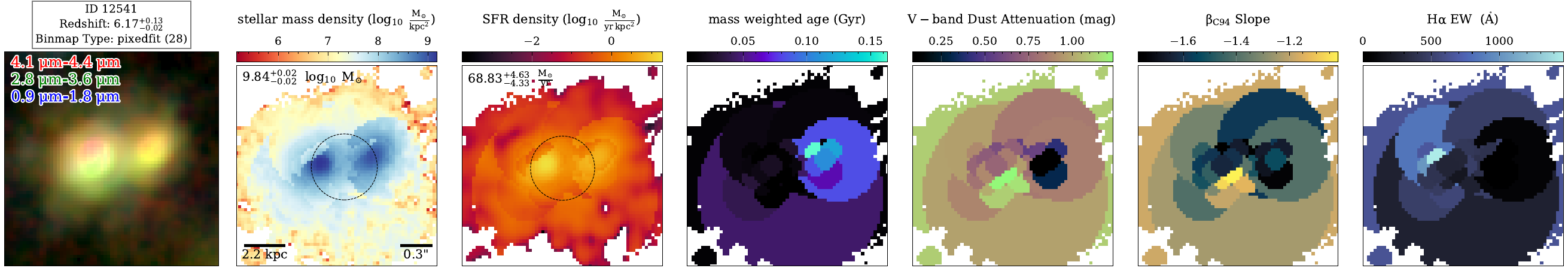}

    \caption{Property maps derived from Bagpipes SED fitting showing stellar mass, SFR, mass-weighted age, dust attenuation, UV $\beta$ slope, and either inferred H$\alpha$ equivalent width or M$_{\rm UV}$, for 7 example galaxies covering a range of redshifts and masses in our sample. On the left we show a representative RGB cutout of the galaxy scaled according to the prescription of  \protect\cite{lupton2004preparing}. Physical and angular scales are shown using scalebars, and the redshift, binning methodology and stellar mass/SFR are given for each galaxy. The top two cutouts and property maps displayed are for the same galaxy with two different binning methods demonstrating the variance in derived parameters with the choice of binning.}
    \label{fig:sed_fits}
\end{figure*}

\autoref{fig:sed_fits} shows the derived property maps for 7 example galaxies, based on our resolved \bagpipes{} SED fitting using a constant SFH, and one of the binning methods (labelled), along with RGB cutouts. We show the half-mass and half-SFR regions, centered on the center-of-mass and center-of-SFR respectively. We also show maps of mass-weighted stellar age. 
The range of inferred dust attenuation can vary by 1-2 magnitudes over a single galaxy, potentially suggesting localised dust production or destruction mechanisms. 

We also show maps of either H$\alpha$ equivalent width, or absolute UV magnitude, depending on whether H$\alpha$ falls within the wavelength range of the available photometry ($z\leq 6.6$). The top two rows show the same galaxy with two of the binning methods, in order to test how reliable overall trends in these properties are to the exact bins used. The smaller bins used in the Voronoi bin maps typically result in a larger spread in pixel values, particularly in properties which are less well-constrained, such as age. Overall trends (e.g. dust, beta slope, age gradients) are generally preserved, although not in all cases.

\begin{figure*}
    \centering
    \includegraphics[width=\textwidth]{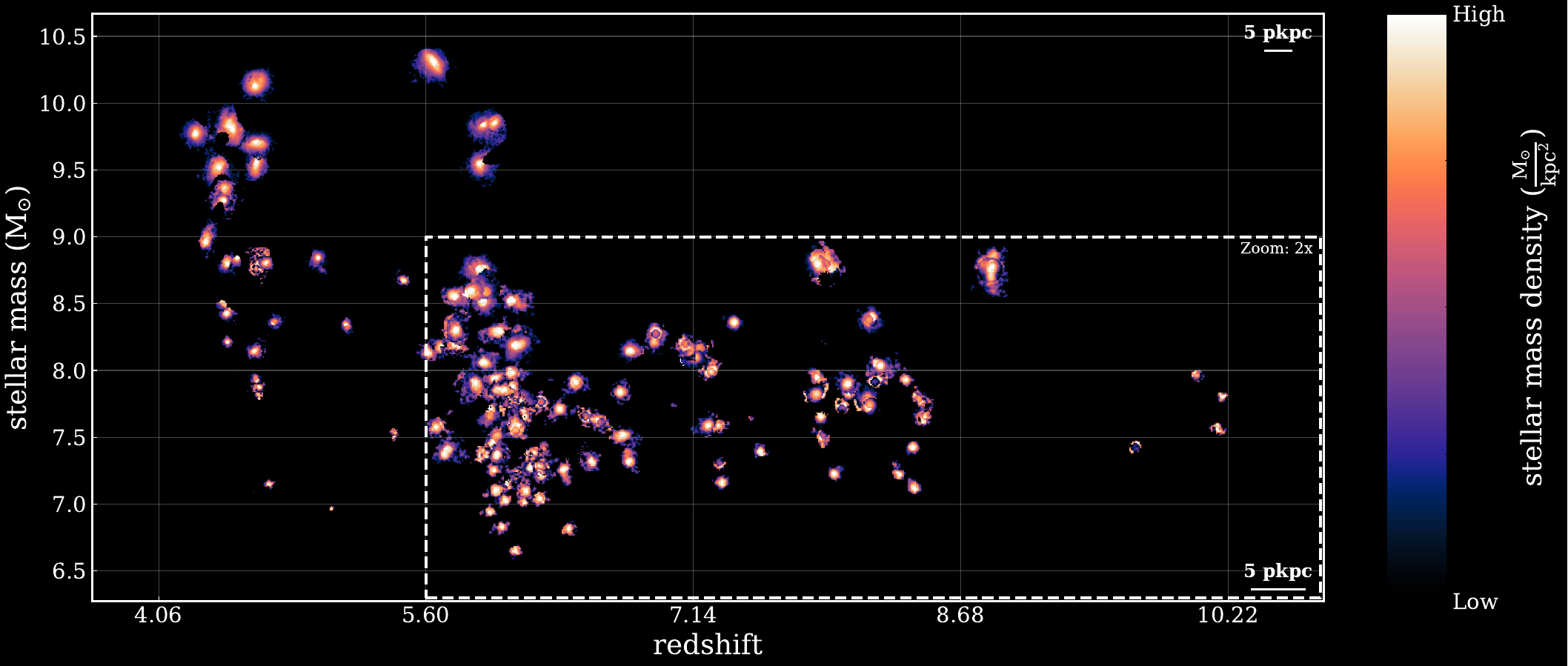}
    \caption{Stellar mass vs redshift for our galaxy sample, where each galaxy is plotted as its stellar mass density map, scaled to the same physical size. Each cutout is scaled independently to preserve detail. Some cutouts may overlap. The indicated section is plotted at 2$\times$ zoom, to improve readability. The binning method used for these maps is the original \pixedfit{} binning.}
    \label{fig:mass_redshift}
\end{figure*}

\autoref{fig:mass_redshift} shows the entire sample on the stellar mass - redshift plane, where each galaxy is represented by its stellar mass density map, scaled to the same physical size. Each map is scaled independently, where yellow shows the regions of highest stellar mass density within each galaxy. Some high-mass galaxy cutouts overlap due to space constraints. Given the large range of galaxies sizes, we have re-scaled an inset region of smaller, higher-redshift galaxies at 2$\times$ the physical scale, in order to improve the readability of the plot.

\subsection{Outshining}
\label{sec:outshining}

\begin{figure*}
    \centering
    \includegraphics[width=\textwidth]{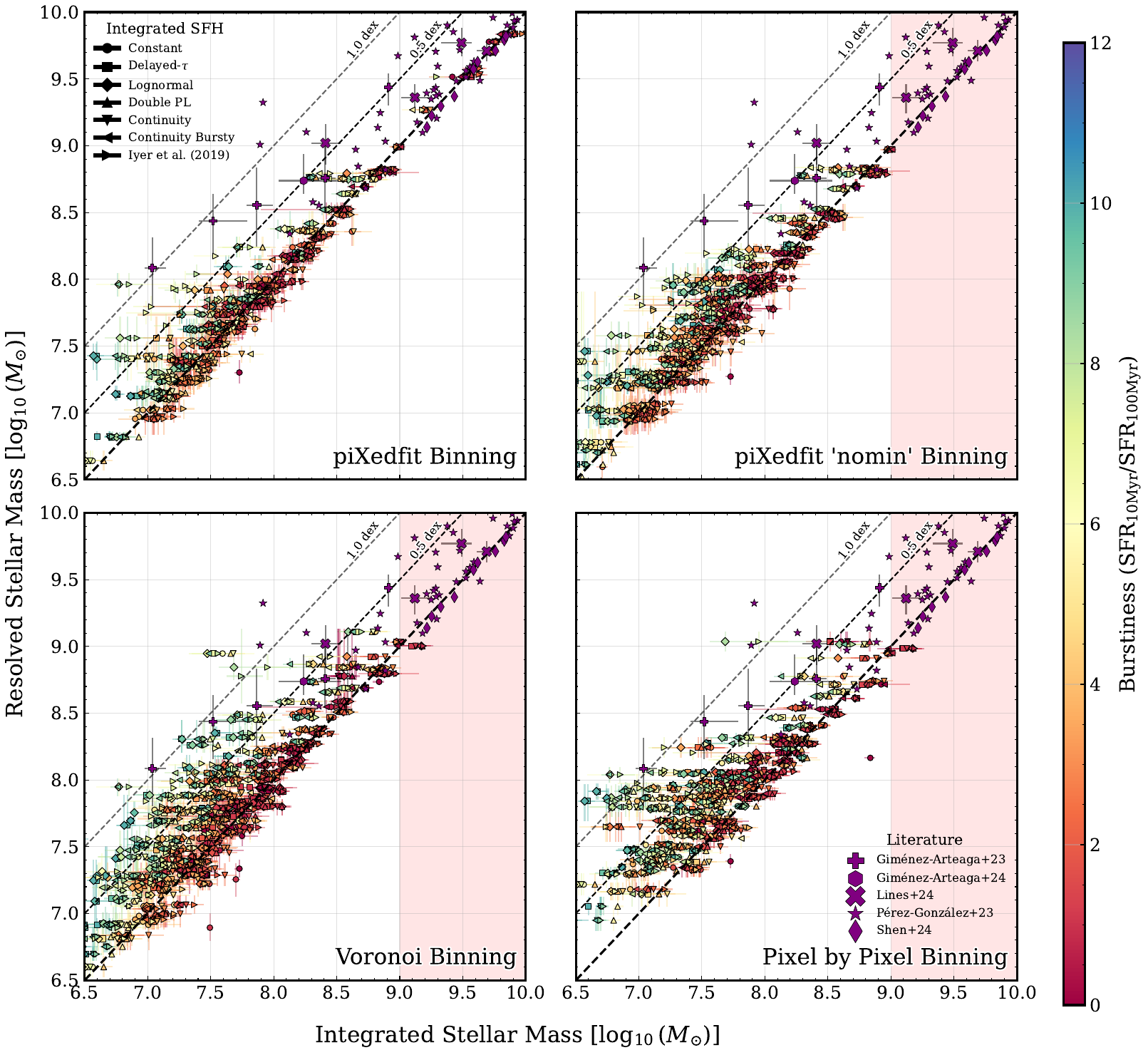}
    \caption{Resolved stellar mass (sum of all individual fitted bins) vs integrated stellar mass (from fit of total galaxy light) for different assumed SFHs, coloured by`burstiness', defined as the ratio of average SFR in the last 10 Myr divided by average SFR in the last 100 Myr. Different plots compare resolved stellar mass estimates using different binning techniques. Shaded red regions indicate the upper mass limit of 10$^{9}$M$_{\odot}$ above which we do not perform SED fitting using the `pixedfit nomin' or `Voronoi' bins, due to the agreement of the resolved `pixedfit' stellar masses with the integrated SFH fits. We show a comparison to comparable outshining results at high-redshift from  \protect\cite{gimenez-artega2022, giménezarteaga2024outshining,perezgonzalez2022, 2024arXiv240910963L} and \protect\cite{2024ApJ...963L..49S} in purple, which we have corrected for magnification where appropriate.}
    \label{fig:mass-comparison}
\end{figure*}

\begin{figure*}
    \centering
    \includegraphics[width=0.8\textwidth]{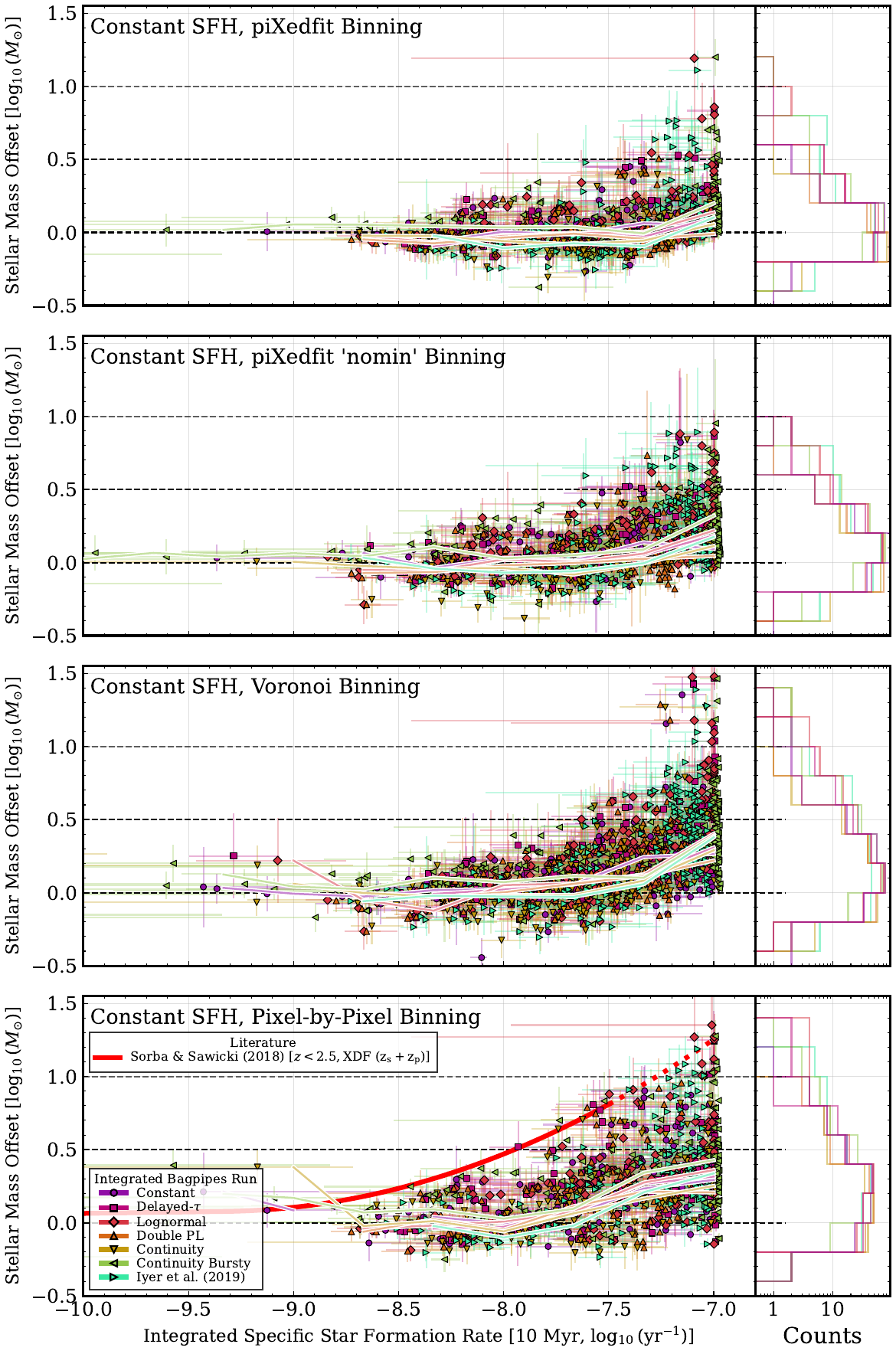}
    \caption{Stellar mass discrepancy ($\Delta$M$_\star=\log_{10}\rm M_{\star, Resolved } - \log_{10}\rm M_{\star, Integrated}$) as a function of integrated sSFR measured on a 10 Myr timescale. Our four different binning methodologies are shown on separate axes. For each binning methodology we show the results for all of our integrated \bagpipes{} models, which we separate by marker color and shape. For each model we also show a median trend line, which typically shows an increasing offset with higher sSFR. On the right of each axes we show a histogram of stellar mass offsets. On the bottom axis we compare to the results of \protect\cite{2018MNRAS.476.1532S}, specifically the results for their combined photometric and spectroscopic sample of XDF galaxies at $z\leq2$ with HST. Our median trend lines show significant lower mass offsets than their best-fit piecewise linear+parabolic fit, suggesting that outshining has less of an impact at high redshift that observed in their sample, or that due to the compact nature of high-redshift higher resolution is needed to look for unresolved outshining.}
    \label{fig:outshining_ssfr}
\end{figure*}

\begin{figure*}
    \centering
    \includegraphics[width=\textwidth]{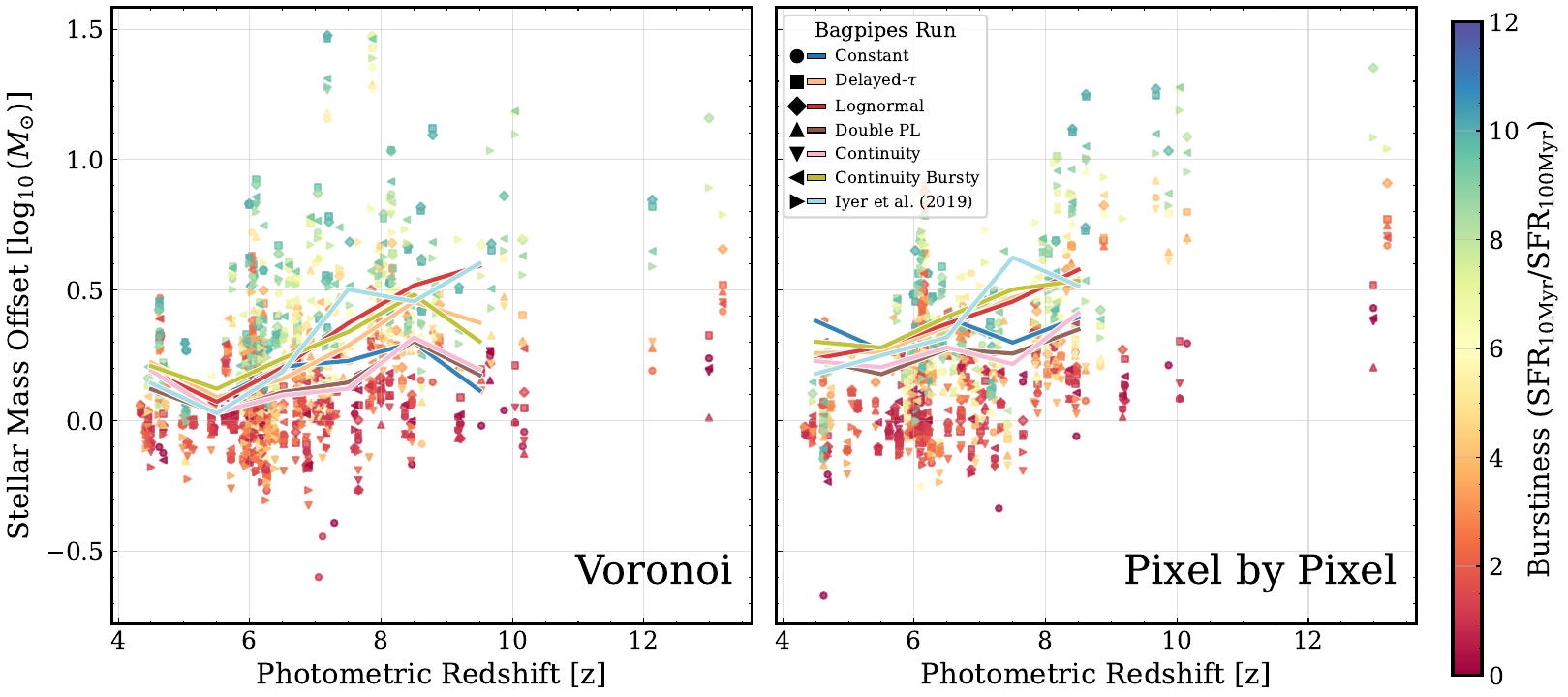}
    \caption{Stellar mass discrepancy ($\Delta$M$_\star=\log_{10}\rm M_{\star, Resolved } - \log_{10}\rm M_{\star, Integrated}$) as a function of photometric redshift for our different integrated SFH models and two of our resolved binning methods. A binned median trend line is shown for each integrated SFH model. We can observe a trend of increasing $\Delta$M$_\star$ with increasing redshift for all SFH models, such that galaxies at higher redshifts in our sample appear to have an increasingly underestimated integrated stellar mass when compared to the resolved fit. This may be because galaxies at higher redshifts are intrinsically more bursty and have bright young stellar populations, or partially due to the  selection effect typical of all photometric surveys which means we are biased towards brighter galaxies with increasing redshift which are more likely to have undergone a recent burst in SFH, but may not be a representative sample.}
    \label{fig:offset_vs_z}
\end{figure*}

By comparing the resolved and integrated \bagpipes{} stellar mass and ages we can look for evidence of outshining, driven by the potential inflexibility of a simple integrated model to explain the complexities and variation in SFH and dust attenuation across a galaxy. We define `outshining' here as any observed stellar mass offset seen between an integrated and resolved stellar mass estimate, such that $\Delta$M$_\star=\log_{10}\rm M_{\star, Resolved } - \log_{10}\rm M_{\star, Integrated}$). 

We calculate total masses for our resolved galaxies by repeating the process of drawing a sample from the posterior stellar mass distribution for each fitted bin and summing them, resulting in a distribution of total summed stellar mass from which we take the median as the stellar mass, deriving uncertainties from the 16th and 84th percentiles. 

We note that we have different numbers of galaxies in each binning sample; from our parent sample discussed in \autoref{sec:sample_selection}, a galaxy is included in a binned sample if the binning criteria segments the galaxy into at least two bins. As the binning criteria (in terms of required SNR) vary between our criteria, the `pixedfit nomin' and Voronoi binned samples contain more galaxies than the `pixedfit' sample. The number of galaxies in each sample is given in \autoref{tab:bin_stats}.

\autoref{fig:mass-comparison} shows a comparison of the stellar masses between each of our  \bagpipes{} integrated SFH models, and the resolved model, for our four different binning methodologies. The number of bins per galaxy increases from top left to  bottom-right. We colour each point by the `burstiness', which we define as the ratio of the 
SFR in the last 10 Myr to the SFR over the last 100 Myr. We can see a correlation between the mass discrepancy and the burstiness in many cases, where galaxies which appear to be undergoing more recent SFR bursts show a larger discrepancy between resolved and integrated masses. The impact of outshining is apparent with all binning methodologies, but the number of galaxies impacted and the observed mass discrepancy changes with the binning methodology.

With our initial \pixedfit{} binning, for galaxies above 10$^9$ M$_\odot$, we do not observe much significant outshining and find generally consistent masses between the resolved and integrated bins. This agrees with the results of \cite{2024arXiv240910963L} and mostly with \citep{perezgonzalez2022}, although they do observe some galaxies in this mass regime with a $>0.5$ dex mass offset. For this reason we chose not to fit these galaxies with the other binning methodologies, as their size and SNR means they are extremely computationally intensive to fit, as they are each segmented into thousands of bins. 

The inferred mass discrepancy varies with the binning methodology and integrated star formation history model used, so we summarise the results in \autoref{tab:outshining_table}. We list the median, mean and max stellar mass offsets between each resolved binning model and integrated SFH history. With the Voronoi binning methodology, we see a few individual galaxies with $>1$ dex discrepancies between the resolved and integrated stellar masses. However when we consider the median offsets across all galaxies in our sample, the median offset even for the most-resolved case does not exceed 0.3 dex, which suggests that whilst outshining can have a large impact on individual galaxies, the overall systematic impact is not that large and is comparable to the potential offsets induced by other SED fitting assumptions such as dust law, parameter priors, IMF or SPS model \citep{2022A&A...662A..86P, wang2024quantifying,2025ApJ...978...89H}. 

\autoref{fig:outshining_ssfr} shows the stellar mass discrepancy between the resolved and integrated masses as a function of sSFR. The discrepancy can be seen to increase with increasing sSFR - i.e. galaxies that have more star formation at a given mass are more likely to show evidence of stellar mass discrepancy driven by outshining. We plot the median trends with sSFR for each \bagpipes{} model. It is apparent that many of the galaxies are located at the sSFR limit of $10^{-7}$ yr$^{-1}$, given the SF timescale of 10 Myr used, and the vast majority of our galaxies have high sSFRs with only a small fraction below $10^{-9}$ yr$^{-1}$. The trend with sSFR does however appear to have a large scatter, and we do not observe a clear departure from the 1:1 relation even for galaxies at the sSFR limit. 

\autoref{fig:offset_vs_z} shows the stellar mass discrepancy as a function of photometric redshift, along with median trend lines for each of our integrated \bagpipes{} SFH fits. We observe a common trend with all integrated SFH models such that the stellar mass offset appears to correlate with redshift, such that galaxies towards higher redshifts have higher stellar mass offsets. In particular, galaxies at $z>8$ appear more likely to have a stellar mass offset of $> 0.5$ dex than those at lower redshift, particularly in our pixel-by-pixel fitting. We do not extend our trend lines across our full redshift range as we have very few galaxies in our sample at $z>10$, and these galaxies are unlikely to be representative of the galaxy population at these redshifts, as we are sampling only the brightest galaxies which have high enough SNR to be analysed in a resolved manner. It is therefore difficult to conclude whether there is an intrinsically different SFH for these galaxies which is less well-fit by normal SFH models and therefore more susceptible to outshining, or if this is just a selection effect, as the brightest galaxies at a given redshift are likely to be those which have undergone a recent burst of star formation, and have higher sSFRs. Size and structure evolution may also play a role; many high-z galaxies are reasonably compact, but resolvable into individual clumps, which may have distinct star formation and metallicity histories \citep{vanzella2023jwst,2024arXiv240208696M, adamo2024bound}.

\begin{table*}
\centering
\caption{Median, mean, standard deviation and maximum of stellar mass offsets ($\Delta$M$_\star=\log_{10}\rm M_{\star, Resolved } - \log_{10}\rm M_{\star Integrated}$) between resolved and integrated \bagpipes{} SED fitting results for different integrated SFH models and resolved binning methods. A positive (negative) $\Delta \rm M$ indicates a higher (lower) resolved than integrated stellar mass. Only galaxies with a resolved stellar mass of M$_* < 10^9 $M$_{\odot}$ are included. All resolved models shown here are fit with the constant SFH model for each bin. Note that the stellar mass offset correlates with sSFR, stellar mass and redshift, which is not represented in these population averages.}
\addtolength{\tabcolsep}{-0.4em}
\begin{tabular}{lcccccccccccccccc}
\begin{tabular}[c]{@{}l@{}}\textbf{Integrated Model}\end{tabular}&\multicolumn{16}{c}{\textbf{Resolved Binning Models}} \\
&\multicolumn{4}{c}{\tt piXedfit}&\multicolumn{4}{c}{\tt piXedfit nomin}&\multicolumn{4}{c}{\tt Voronoi}&\multicolumn{4}{c}{\tt pixel-by-pixel} \\
&Mean&Median&Std. Dev.&Max&Mean&Median&Std. Dev.&Max&Mean&Median&Std. Dev.&Max&Mean&Median&Std. Dev.&Max \\
 \hline
Constant SFH &0.02&0.06&0.12&0.45&0.07&0.14&0.16&0.82&0.16&0.25&0.23&1.16&0.23&0.33&0.24&1.00 \\
Delayed SFH &0.03&0.11&0.17&0.83&0.10&0.19&0.20&0.87&0.15&0.35&0.28&1.46&0.26&0.40&0.28&1.25 \\
Lognormal SFH &0.06&0.20&0.22&1.19&0.12&0.23&0.22&0.89&0.20&0.40&0.30&1.48&0.27&0.46&0.30&1.27 \\
Double power-law SFH &-0.03&0.02&0.13&0.51&0.01&0.08&0.15&0.73&0.09&0.20&0.21&1.18&0.18&0.27&0.23&0.96 \\
Continuity SFH &-0.03&0.01&0.13&0.40&0.02&0.08&0.17&0.84&0.08&0.20&0.23&1.27&0.15&0.26&0.25&0.99 \\
Continuity Bursty SFH &0.09&0.20&0.20&1.20&0.14&0.23&0.20&0.95&0.22&0.36&0.26&1.31&0.30&0.43&0.27&1.28 \\
\cite{Iyer2019NonparametricFormation} SFH &0.03&0.20&0.26&1.11&0.07&0.23&0.25&0.89&0.23&0.36&0.29&1.28&0.26&0.43&0.31&1.19 \\

\end{tabular}
\label{tab:outshining_table}
\end{table*}

It is clear from these results that outshining can have a large impact on individual galaxies, and care should be taken when choosing SFH priors for fitting high-z galaxies, particularly those with indications of recent star formation. 

\subsection{Inferred Stellar Population Ages and SFHs}

\begin{figure}
    \centering
    \includegraphics[width=\columnwidth]{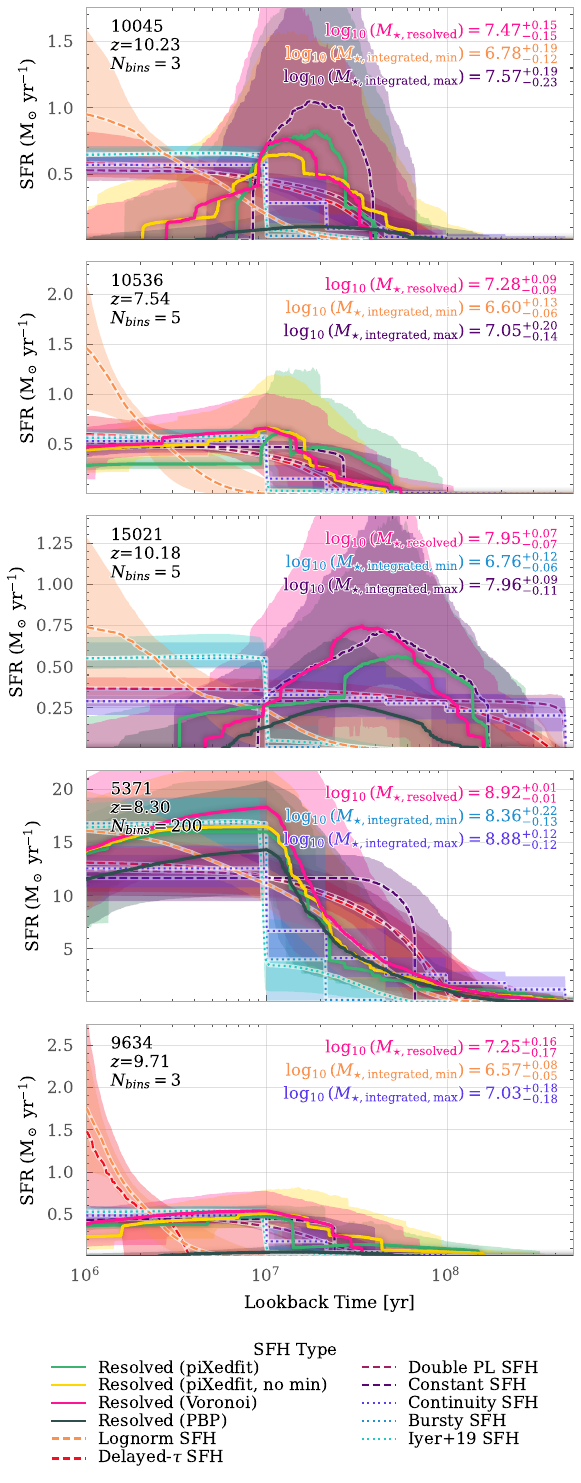}
    \caption{Example SFHs of galaxies with >0.5 dex of outshining. Galaxy ID and redshift are labelled on each axis. The total surviving stellar mass from the resolved fitting using the Voronoi binning is shown, as are the maximum and minimum stellar masses for our integrated models with different SFH parametrizations. }
    \label{fig:outshining_sfhs}
\end{figure}

When we compare the inferred mass-weighted ages between the resolved and integrated fits we see mixed results. The integrated constant and continuity bursty SFHs typically have significantly younger stellar ages than the resolved constant case. The non-parametric `continuity' SFH however often has older mass-weighted ages than the resolved constant SFH model, which may just be a consequence of the parametrization of the model and the size and width of the SFH time bins used. In a few cases the lognormal and delayed exponential star formation history models favour gradually rising star formation histories, with older ages than the resolved case, but the majority of the time they favour very young ages, in some cases essentially instantaneous bursts within the last 1 - 3 Myr, whereas the resolved case will suggest ages of 10's to 100's of Myr. 

\autoref{fig:outshining_sfhs} shows example star formation histories for 6 galaxies with $>0.5$ dex of of mass bias induced by suspected outshining (when compared to one or more resolved SFHs), for both the integrated and resolved fits. The integrated fits, and particularly the lognormal, delayed exponential and constant SFH models are typically fit as recent rising SFH models, with no older stellar populations which are more commonly seen in the resolved and non-parametric cases. We can also compare the recovered star formation histories for our different resolved binning models, and see that in some cases the resolved SFH estimates look similar for all binning methods, whilst others show significant differences when the galaxy is split into a different number of bins.

\subsection{Emission Line Tracers and Outshining}

As discussed in \autoref{sec:emission_line_properties}, we calculate global and integrated equivalent widths for the H$\alpha$ line for the galaxies in our sample at $z \leq 6.6$ using the photometric excess between overlapping medium/wide NIRCam bands. H$\alpha$ emission traces recent SFR ($\sim 10$ Myr timescale). Given that we see a correlation between `burstiness' and sSFR and the amount of outshining observed, we might expect to also observe a correlation with H$\alpha$ EW. 

\begin{figure}
    \centering
    \includegraphics[width=\columnwidth]{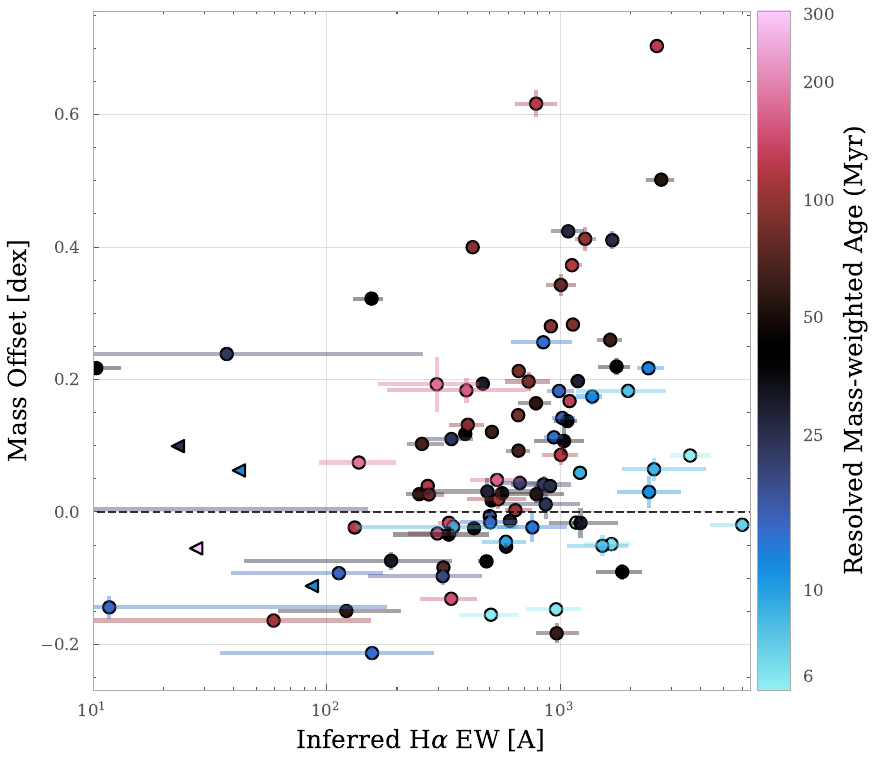}
    \caption{Inferred H$\alpha$ equivalent width from a photometric excess, vs mass difference between Voronoi binned resolved SFH model and the double power-law integrated SFH model. Points are coloured by resolved mass-weighted age, and only shown for galaxies at $z\leq$ 6.6 where we can directly observe H$\alpha$ emission. Strong H$\alpha$ emission indicates recent star formation (within the last 10 Myr). Points marked with a triangle are an upper limit, as no H$\alpha$ emission is inferred given the uncertainties. We observe a correlation between the outshining-induced mass discrepancy and the H$\alpha$ EW, such that higher EW galaxies typically have , but there is considerable scatter among individual galaxies. }
    \label{fig:ew}
\end{figure}

\autoref{fig:ew} shows a comparison between the stellar mass discrepancy of our resolved SFH model using Voronoi binning and the integrated double power-law model with inferred H$\alpha$ equivalent width. The overall relationship seen is relatively consistent for different choices of integrated/resolved model. We see a correlation between mass offset and EW, such that galaxies with the largest stellar mass offsets have higher EWs ($\approx 1000$\AA{}), but with a significant amount of scatter. Galaxies with low EW or no inferred H$\alpha$ emission appear to have stellar masses consistent within 0.2 dex in the resolved and integrated cases, but there are also a few cases of galaxies with very high equivalent widths ($\geq$ 2000\AA{}) with consistent stellar masses, which occurs when the resolved and integrated SFH models both favor a short formation timescale, as can be seen from the color-scale showing the resolved mass-weighted age. Typically those galaxies with high EWs which have significant mass offsets have longer resolved mass-weighted ages ($\geq$ 100 Myr), which aren't accurately recovered with the integrated SED fitting.

\subsection{Comparison of Integrated SFH Models}

\begin{figure}
    \includegraphics[width=\columnwidth]{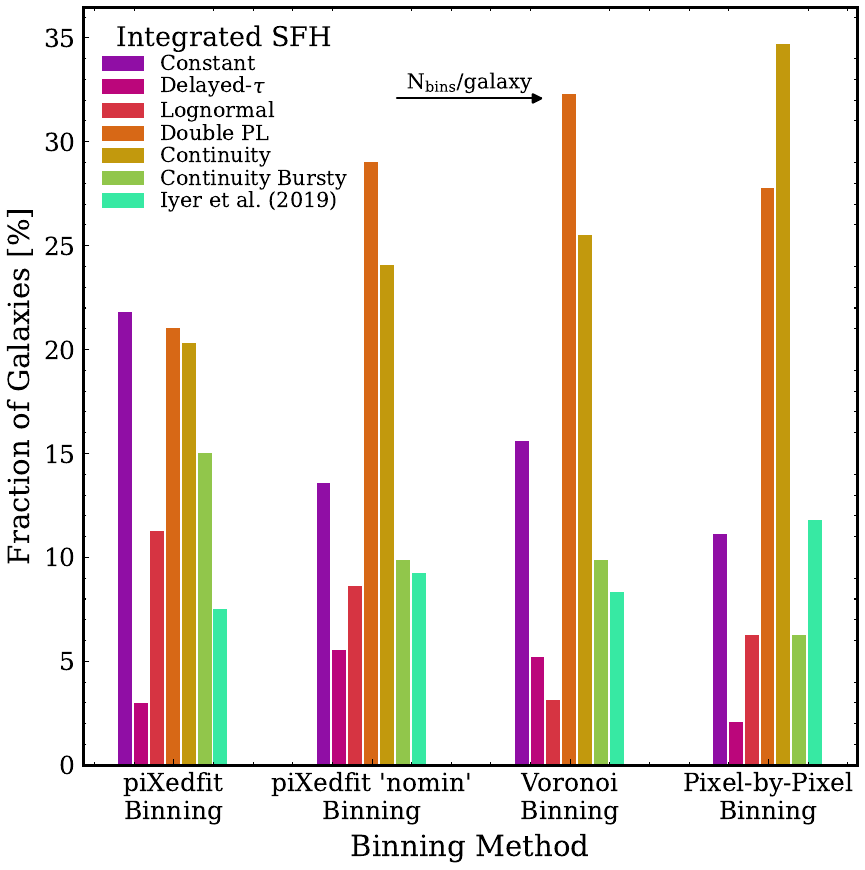}
    \caption{Bar chart showing which integrated SFH model has the lowest $|\Delta$M$_\star|=|\log_{10}\rm M_{\star, Resolved } - \log_{10}\rm M_{\star, Integrated}$)| for each of our binning methodologies. The number of bins per galaxy increases from left to right. The `continuity' model of \protect\cite{leja2019measure, tachella2022} and the double power-law model produce the closest integrated stellar masses to our resolved estimates in the plurality of cases.}
    \label{fig:best_int_sfh}
\end{figure}

One of the purposes of testing multiple forms of integrated SFH, with different prior assumptions on the timescales and shapes of SFH is to determine which integrated SFH produces the closest stellar mass estimates to our resolved fits. We calculate the mass discrepancy for each galaxy between all integrated SFH models and each resolved resolved binning methodology. We then take the least-discrepant model (smallest $\Delta$M$_\star|=|\log_{10}\rm M_{\star, Resolved } - \log_{10}\rm M_{\star, Integrated}$)| as the best SFH model for that galaxy with that resolved binning method. The results of this analysis are shown in a bar chart in \autoref{fig:best_int_sfh}, which shows the fraction of each integrated SFH model produced the closest model to each resolved estimate.

The relative performance of each integrated model changes slightly with the binning method, but we see similar overall trends. With the `pixedfit' binning, the constant star formation history wins a narrow plurality over the non-parametric `continuity' model and the double power-law model. This may be because this is is our most conservative binning case, which attempts to respect correlations smaller than the PSF and has the largest SNR requirements per bin, resulting in the fewest number of bins per galaxy. This means a significant fraction of the galaxies in this binning scenario are only fit by 2-4 bins, which may result in similar fits to the integrated constant SFH fit, particularly if one of the bins dominates over the others. Whilst this may suggest the performance of this simple model in this case is an artefact of our choice of resolved SFH, we also note that this binning is the only one we fit to the most massive galaxies ($\geq 10^9$M$_\star$), which may be better fit by the constant SFH model. 

When we look at our other binning methods with larger numbers of bins per galaxy, we typically see two models dominate over the rest. The models with the best performance here are the double power-law SFH model and the non-parametric continuity model of \cite{leja2019measure}. However all tested integrated SFH models produce the closest stellar mass estimates for some galaxies, so it is not a case of one model always outperforming the other models. 

Both the double power-law and continuity models are reported in the literature to produce systematically higher stellar masses than other assumed forms of the SFH when fitting integrated photometry \citep{leja2019measure,tacchella2022stellar}. These systematically higher masses are typically found because these model priors encourage older stellar populations and a more gradual build-up of stellar mass. Whilst these models produce the closest stellar masses to the resolved estimates, they are still systematically offset at high sSFR and do not fully recover the complexities of the resolved SFHs. 

It is worth noting that these models were first introduced to fit the SFHs of galaxies at lower redshifts, which have significantly longer timescales to form their stellar mass, in order to allow longer star-formation histories than found with typically used star-formation histories such as the exponential SFH model. In particular the double power-law model has often been used to fit quiescent galaxies with old stellar populations \citep[e.g.][]{carnall2019vandels}. So whilst these models appear to be producing overall stellar mass estimates closer to our resolved measurements, it is important to consider whether these models can produce reasonable star-formation histories at the highest redshifts. Simulations suggest that stochastic star formation histories are common in low-mass galaxies as they have shallow potential wells and are sensitive to star-formation induced feedback, which are often averaged over as smooth SFHs when SED fitting, leading to scatter in stellar mass estimates due to the enforced shape of the SFH prior \citep{iyer2020diversity,pallottini2023stochastic}. Models that allow rapid changes in SFR have been found to work well and produce more accurate masses when compared to sims, which agrees with our results \citep[e.g.][]{lower2020well,cochrane2024highzstellarmassesrecovered}.

As well as comparing which model produces the closest mass estimates across all cases, we consider the relative behaviour of different integrated SFH models as a function of sSFR, as shown in \autoref{fig:outshining_ssfr}. We see similar overall trends, in that the median trend line shows an increasing mass-offset between integrated and resolved stellar mass with increasing sSFR, but we can see slight individual differences between the different models. For example, with our pixel-by-pixel model we can observe that the continuity bursty model shows a larger average offset for galaxies with high sSFRs ($\sim 10^{-7}$ yr$^{-1}$) than the other models. The observed trends vary with the binning method, but we see the double power-law and continuity SFH models typically produce the closest mass estimates even at high sSFR. 

Overall, the star formation history models which can lead to the largest offsets in individual cases are typically those which, if the SFH is inferred as a very recent burst (see examples of delayed$-\tau$ or lognormal SFH fits in \autoref{fig:outshining_sfhs}), have no flexibility to also infer any older stellar population. A rapidly rising lognormal or delayed$-\tau$ SFH for galaxies with high ongoing star formation often has all the stellar mass in the galaxy formed in the last $<10$ Myr, with no flexibility to also include an older stellar population without lowering the current SFR or assuming a much flatter SFH. This is why `burstiness' scales strongly with the inferred mass-offset, as most of these parametric star formation histories do not allow for old stellar populations alongside a steeply rising burst of recent SFH.

Models which allow multiple periods of SFH and rapid changes in SFR can allow for young and old stellar populations to be more easily inferred simultaneously, but our results show that there is still potential room to improve the SFH priors we assume to fully capture the complex SFHs of high-redshift galaxies. 

\subsection{Comparison to other studies}
\label{sec:comparison}
\autoref{fig:mass-comparison} shows the results of previous studies of outshining from \cite{gimenez-artega2022, giménezarteaga2024outshining, perezgonzalez2022, 2024arXiv240910963L} and \cite{2024ApJ...963L..49S}. These studies have studied the effect of outshining on galaxies across different redshift and mass regimes. 

We compare our findings to studies by \cite{gimenez-artega2022} and \cite{giménezarteaga2024outshining}, (hereafter GA23 and GA24), which examined six galaxies with stellar masses between M$_{\star}=10^8$-$10^{9.5}~\rm{M}_\odot$. For their five moderately lensed galaxies ($\mu < 3$), GA23 found stellar mass offsets of 0.5-1 dex as shown on \autoref{fig:mass-comparison}. GA24 studied one strongly lensed galaxy ($\mu = 29$) with a de-lensed mass of $\approx10^{8.7}~\rm{M}_\odot$ at $z=6$, finding a 0.5 dex stellar mass discrepancy between resolved and integrated analyses.

Our results align with theirs when using Voronoi or pixel-by-pixel binning, showing similar or greater mass discrepancies. However, their most massive galaxy (ID 8140) shows a larger discrepancy than any comparable galaxy in our sample, possibly due to differences in SFH or resolution.

Our study improves upon previous work by using 19 photometric filters (versus their 5-6), covering a wavelength range of 0.35-5$\mu$m and including 7 medium-band filters. This allows better differentiation between emission lines and continuum features, improving constraints on recent SFH and dust attenuation. Our larger sample ($>$200 galaxies) suggests the extreme outshining observed by GA23 and GA24 is not representative of typical galaxies at these redshifts. 

GA24 tested multiple forms of the integrated SFH to determine which model produces stellar mass estimates closest to their resolved fit, finding that the double power-law fit produced the closest stellar mass estimate, but they also note that the fit could not reproduce the extreme emission line equivalent widths observed in their highly lensed galaxy with NIRSpec ($\geq 3000$\AA{}). We also find that the double power-law model performs well when comparing resolved and integrated masses as it typically infers higher stellar mass estimates than other integrated SFH models. 

At higher masses we compare to \cite{2024arXiv240910963L}, who studied 4 star-forming galaxies at $z\approx 5$ from PRIMER with stellar masses of M$_{\star}\sim10^{9-9.5}$ M$\odot$, finding typically consistent integrated and resolved stellar mass estimates, apart from for their lowest mass galaxy, for which they found a 0.5 dex offset in stellar mass. As outshining is thought to be a mass-dependent effect, with larger mass offsets at lower stellar mass, their observation of consistent resolved and integrated stellar masses is not unexpected and agrees with our results in this mass regime with our \pixedfit{} binning results. They test a number of integrated SFH models, but we show their delayed exponential SFH points in \autoref{fig:mass-comparison}. 

We also show a sample of star-forming galaxies at lower redshift ($0.2 \leq z \leq 2.5$) from \cite{2024ApJ...963L..49S}, with masses M$_{\star}\geq 10^{9}$ M$_\odot$, who compare integrated and resolved masses using a delayed exponential SFH. This sample of low-z galaxies are consistent with no observed outshining in this higher mass regime. 

We also plot the sample of \cite{perezgonzalez2022}, who analysed red galaxies at $2 \leq z \leq 6$. These galaxies are primarily consistent with the 1:1 relation for resolved and integrated masses, particularly at M$_\star > 10^{9.5}$M$_\odot$, but some individual objects have more than 1~dex offsets, which we also see in a few cases. 

At $z \leq 2.5$ we can qualitatively compare with the results of \cite{2018MNRAS.476.1532S}, who also looked at the systematic effects of outshining by performing resolved pixel-by-pixel fits of galaxies in the Hubble XDF using HST ACS/WFC and WFC3IR observations. They found a broken power-law shape for the relationship between integrated sSFR and stellar mass-offset, which we show for their combined photometric and spectroscopic sample in red on \autoref{fig:outshining_ssfr}, where all galaxies above an sSFR of $\sim10^{-8}$ yr$^{-1}$ are offset from the 1:1 relation. We do not observe such a dramatic turn away from the line, and indeed many of our galaxies even at the sSFR limit of $10^{-7}$ yr$^{-1}$ are consistent with no stellar mass offset between integrated and resolved estimates. The reason for the smaller impact of outshining we infer compared to their results is not clear, but there may be unresolved outshining within our sample that we can not distinguish with the PSF and resolution of NIRCam, or the larger number of photometric bands may allow our integrated SED fitting to perform better in most cases, resulting in a lower overall systematic effect.
Another consideration is the maximal age of a obscured stellar population is significantly shorter for galaxies in our sample than those of \cite{2018MNRAS.476.1532S}. Assuming star-formation could start no earlier than $z=20$, the maximally old stellar population for a $z=5$ galaxy in our sample is $\leq 1 \; {\rm Gyr}$ and indeed at $z=10$ it is $\leq 300$ Myr. For \cite{2018MNRAS.476.1532S}, they consider galaxies at $z\leq$2.5, and so the age of a maximally old stellar population is anywhere from 2.5 Gyr to $\geq$10 Gyr. Whilst outshining of a significant stellar population this extreme is unlikely, this serves to illustrate that the systematic effects of outshining may be more apparent at lower-redshift, even if we see individual cases of extreme stellar mass offsets at high-$z$.

Overall for our results with comparable binning to these observational studies at similar redshifts (the pixel by pixel and Voronoi cases) we see consistent amounts of outshining among the most discrepant galaxies in our sample, but the overall impact of outshining across the full sample (at scales which we can distinguish) appears lower than observed at $z<2.5$, which may be partially due to NIRCam medium-band observations enabling better SFH constraints in a larger fraction of the sample when compared to previous studies with fewer photometric bands or pre-JWST measurements from HST+Spitzer.

The effect of outshining has not been studied in many cosmological simulations, but \cite{2024ApJ...961...73N} looked at the effect of outshining in two numerical simulations with different feedback models. They demonstrated that bursty star formation can lead to outshining, resulting in systematic underestimation of stellar masses by nearly 1 dex. They compare different integrated SFH models to the true SFH, including some of the same parametric and non-parametric models we have tested, and find that models which allow rising SFH, as well as rapid variations in SFH, perform the best, but that none of the models tested recover the complex SFHs of the simulated galaxies well. However \cite{cochrane2024highzstellarmassesrecovered} showed that recovering the stellar masses of simulated galaxies from the SPHINX20 simulation with \bagpipes{} is broadly accurate (within a factor of 3), but that at low stellar masses ($\leq 10^8$ M$\odot$) SED fitting can in fact systematically overestimate stellar mass estimates. Our results are more consistent with those of \cite{cochrane2024highzstellarmassesrecovered}, in that we find integrated stellar mass estimates are generally reliable, with some systematic effects due to outshining, and not as unreliable as the results of \cite{2024ApJ...961...73N} would suggest.

\subsection{Impact of Binning Criteria and Methodology}

As discussed in \autoref{sec:binning} we test four different methodologies for grouping pixels to increase SNR when SED fitting. These methodologies group pixels on different spatial scales and with different SNR requirements, resulting in differing total number of fitted bins for each galaxy. One of the purposes of testing different binning methodologies is to examine whether full pixel-by-pixel analysis is necessary, or if we can approximate the effect and improve uncertainties in derived parameters using pixel binning. 

We see systematically different amounts of outshining depending on the binning methodology used. For our first binning methodology, \pixedfit{} binning, the stellar mass estimates for the vast majority of galaxies are consistent within 0.5 dex. This is the only binning methodology where we constrain the minimum size of the bins to attempt to minimize the correlation between bins due to the PSF. This typically results in low numbers of bins, which appear not to provide the spatial resolution necessary to observe outshining in most cases. In a typical $10^{7.5}$ M$_\odot$ galaxy using this binning, it will only be split into 2 to 3 bins, with the majority of the light still contained within the central bin, which then has potentially unresolved outshining given a typical bin size of $\approx$ 1 kpc or more. We also observe with this binning methodology some cases where the resolved mass is actually smaller than the integrated mass by up to 0.2~dex, which is most common when comparing the non-parametric `continuity' SFH to our resolved estimate. This is likely to occur in galaxies with little evidence of outshining due to recent star formation, as the non-parametric SFHs are known to produce systematically higher stellar masses than other methods \citep{leja2019measure,tacchella2022stellar, 2025ApJ...978...89H}. 

As we move to our other binning techniques, where each galaxy is split into smaller regions with lower SNR/bin, we see systematically larger stellar mass offsets, as shown in \autoref{tab:outshining_table}. One other possibility is that the decreasing SNR/bin introduces a systematic bias towards higher masses for low SNR regions, which has been suggested by \cite[e.g.][]{gallazzi2009stellar}. This was investigated by \cite{2018MNRAS.476.1532S} who found a weak mass bias with decreasing SNR of only 2.2\%, which was not significant compared to their outshining offset. This may be evidence for further unresolved outshining below the resolution limit of NIRCam, so the offsets seen with our pixel-by-pixel analysis may be lower limits on the true resolved mass. 

Our binning methodology with the most bins was the `Voronoi' binning method, rather than our individual pixel-by-pixel binning method, as in many cases SNR constraints on individual pixels meant that only the centers of low-mass galaxies could be fit, whereas with Voronoi binning we could combine regions of low surface brightness to increase the SNR, whilst still fitting individual pixels above the SNR limit. 

Our results show that the inferred stellar masses, as well as other properties (e.g. those shown in \autoref{fig:sed_fits}) can depend on the binning methodology used, so we urge other studies to carefully consider their choice of binning when conducting a resolved analysis. The different binning methodologies tested here are appropriate for different situations. For massive galaxies, binning methodologies such as the `pixedfit' binning produce similar results to our other binning methodologies with a much smaller computational cost, but for more compact sources at higher redshift we suggest `Voronoi' binning can combine the spatial resolution advantages of pixel-by-pixel analysis with the SNR improvements of binning, as long as care is taken not to overinterpet features below the scale of the PSF. 


\subsection{Do medium-band filters help better constrain integrated SFHs?}

In this analysis we have used HST ACS/WFC and JWST NIRCam imaging of the JADES Origins Field, which includes 19 photometric bands, including 7 NIRCam medium-band filters; F162M, F182M, F210M, F250M, F300M, F335M and F410M which are of comparable depth to the wideband filters (see \autoref{fig:depths_areas}). However in many JWST surveys, particularly those surveying large cosmic volumes, such as COSMOS-Webb \citep{2021jwst.prop.1727K} or PRIMER \citep{2021jwst.prop.1837D}, medium band filters are not available and SED fitting derived estimates of properties such as stellar mass must rely only on wideband photometry. Medium-band photometry allows SED-fitting codes to much more accurately disentangle the stellar continuum and nebular emission lines, whereas with wide-band photometry alone, a Balmer break indicative of an aged stellar population can easily be fit with high equivalent width emission lines, or vice versa, leading to incorrect measurements of both stellar mass and star formation rate \citep{trussler2024epochs, cochrane2024highzstellarmassesrecovered}.  

In this section we consider whether the overall relative consistency of the derived stellar masses from resolved and integrated SED fitting, except in the most extreme cases, is in part due to the availability of medium-band photometry. We note that the previous studies of outshining at high-redshift discussed in \autoref{sec:comparison} have relied primarily on NIRCam wideband observations for both their resolved and integrated analysis, and in some cases only a subset of the available wideband filters. 

In order to test this we refit our integrated photometry for all galaxies including only our ACS/WFC and NIRCam wideband photometry using \bagpipes{}. We test a subset of our \bagpipes{} SFH models, including one parametric (double power-law) and one non-parametric (continuity-bursty) SFH model. We fix the photo-$z$ of the fit to that of our resolved fit in order to consider the impact of the inferred SFH alone, rather than systematic offsets in photo-$z$ accuracy. 

We observe a larger scatter in stellar mass between our resolved measurements (with medium-bands) and our wide-band only integrated fits. The lack of medium band observations mean that some galaxies shift towards higher stellar masses by 0.2-0.4~dex due to the lack of information to constrain the relative strength of emission lines and the continuum, which in some cases actually reduces the stellar mass offset with the resolved fit. However the overall scatter between resolved and integrated stellar masses increases by 0.3 to 1~dex, which demonstrates the power of medium-band observations in improving SED fitting constraints. 

\subsection{Impact of Resolved SED Fitting Model}

Our SED fitting model for each resolved bin, assumes a constant star formation history which switches on (and can switch off) at a given look-back time. For an individual draw of the SFH posterior this results in a flat SFH, but when sampling quantiles from the full posterior a more complex SFH can be obtained. As each bin fits only a subset of the total galaxy light, we can still infer differences in stellar age and sSFR from this simplistic star formation history, but it is possible that the SFH model is not fully capturing the star formation history of each individual bin, particularly in the case of the \pixedfit{} binning, which has the largest bins. In order to test whether the amount of outshining inferred depends strongly on the chosen star formation history for each bin, we refit all resolved \pixedfit{} bins using a flexible star formation history, specifically the non-parametric `bursty' SFH of \cite{tachella2022}. This star formation history infers a star formation history in six pre-determined bins in lookback time starting at $z=20$, and allows for rapid star formation changes between bins, which may more accurately model stochastic star formation. 

\begin{figure*}
    \centering
    \includegraphics[width=\textwidth]{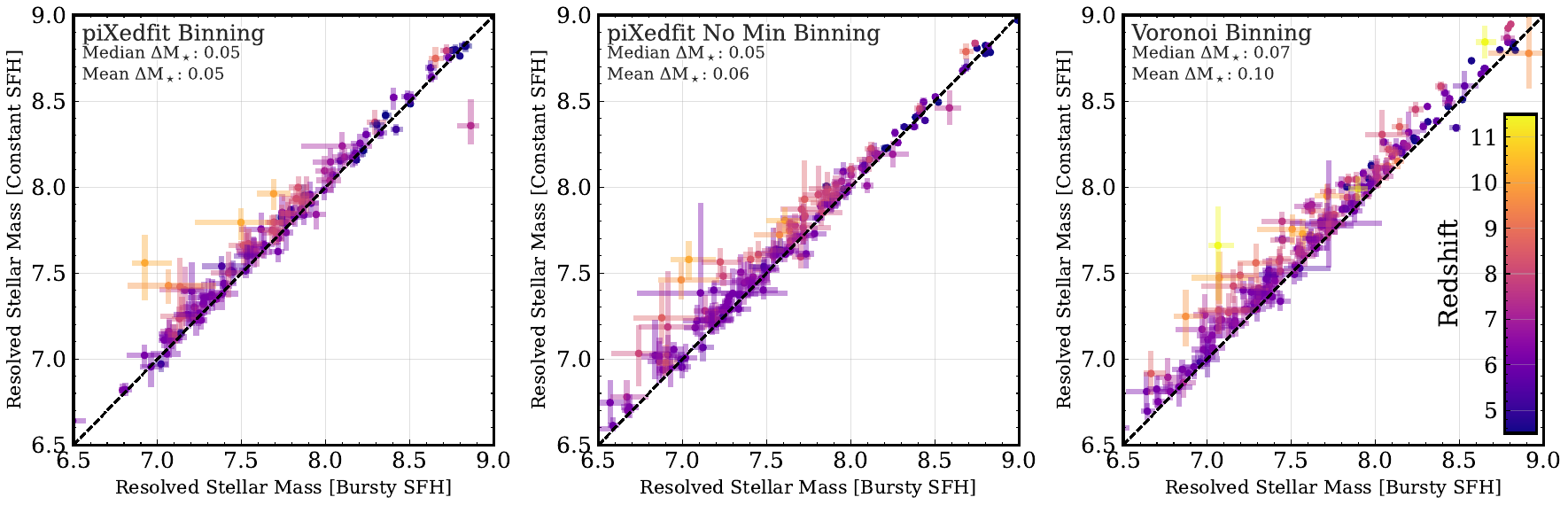}
    \caption{Comparison of the resolved stellar mass estimates using a constant SFH and a non-parametric `continuity bursty' SFH for each pixel/bin. We show this comparison for three of our binning cases. Points are coloured by the redshift, showing systematically slightly lower masses (by up to 0.1 dex on average) when using the more complex non-parametric model.}
    \label{fig:bursty_vs_constant}
\end{figure*}

\autoref{fig:bursty_vs_constant} shows a comparison of the resolved stellar masses derived with the non-parametric `bursty' SFH and our original constant SFH model, which we have split between three of our binning methodologies. We see fairly consistent results for these methodologies and find that resolved masses with the constant SFH are typically systematically offset higher by around 0.05 to 0.1 dex. This means our constant SFH model is actually producing slightly higher stellar mass estimates than the bursty non-parametric model. This may be because for each fitted bin, the bursty SFH model allows for a more rapid decline in SFR with increasing look-back time then the constant SFH model, resulting in lower stellar masses. Overall the effect is subdominant compared to the observed outshining effect, and the effects of outshining are still present for some galaxies if we compare our integrated SED fitting results with the resolved bursty SFH model. 

We do not test the impact of other assumptions, such as dust law, parameter priors/hyperpriors or SPS model as that is beyond the scope of this work, but we explored the variation of these assumptions on integrated SED fitting in \cite{2025ApJ...978...89H}. Other studies which look at systematic effects of SED fitting include \citep{2019ApJS..240....3H, Pacifici2022TheTechniques, leja2019measure, 2022ApJ...935..146S} among many other studies.

\subsection{Choice of SED Fitting Tool}

Our primary analysis has used \bagpipes{}, which is a commonly used SED fitting tool \citep{Carnall2018_Bagpipes, carnall2019measure}. In any study which relies on inferences from SED fitting, it is important to consider whether any results obtained are dependent on the underlying assumptions of the SED fitting tool. One way we test this is to refit the integrated and resolved photometry for all galaxies using \db{}, which is an alternative SED fitting tool. 

\db{} is an SED fitting which combines the FSPS stellar library \citep{conroy2010fsps} with a non-parametric estimate of star formation history using Gaussian Processes \citep{Iyer2019NonparametricFormation}. \cite{Iyer2019NonparametricFormation} shows that complex and varied star formation histories can be recovered from photometry using this methodology. 
We use the same binning methodologies as presented in \autoref{sec:binning}, and fix the redshift during the fitting to the same redshift as our resolved \bagpipes{} analysis. Whilst we also test this parametrization of the star-formation history with \bagpipes{}, \db{} offers an alternative fitting procedure and different underlying assumptions about the stellar templates and isochrones used. 

We chose \db{} specifically to test whether the flexibility of its non-parametric SFH model may allow it to better account for the effect of outshining. Our model uses the \cite{calzetti2000dust} dust law and a \cite{2003PASP..115..763C} IMF. We use uninformative uniform priors on stellar mass ($5 \leq \log_{10} \rm M_\odot/M_\star < 12$), metallicity ($-4 \leq \log_{10} Z_{*}/Z_\odot \leq 0.5)$ and dust attenuation ($0 \leq \rm{A}_{\rm V} \leq 6$), drawing 500,000 realizations to build our template library for fitting. We use the default $\chi^2$ likelihood evaluation for fitting. 

\begin{figure*}
    \centering
    \includegraphics[width=\textwidth]{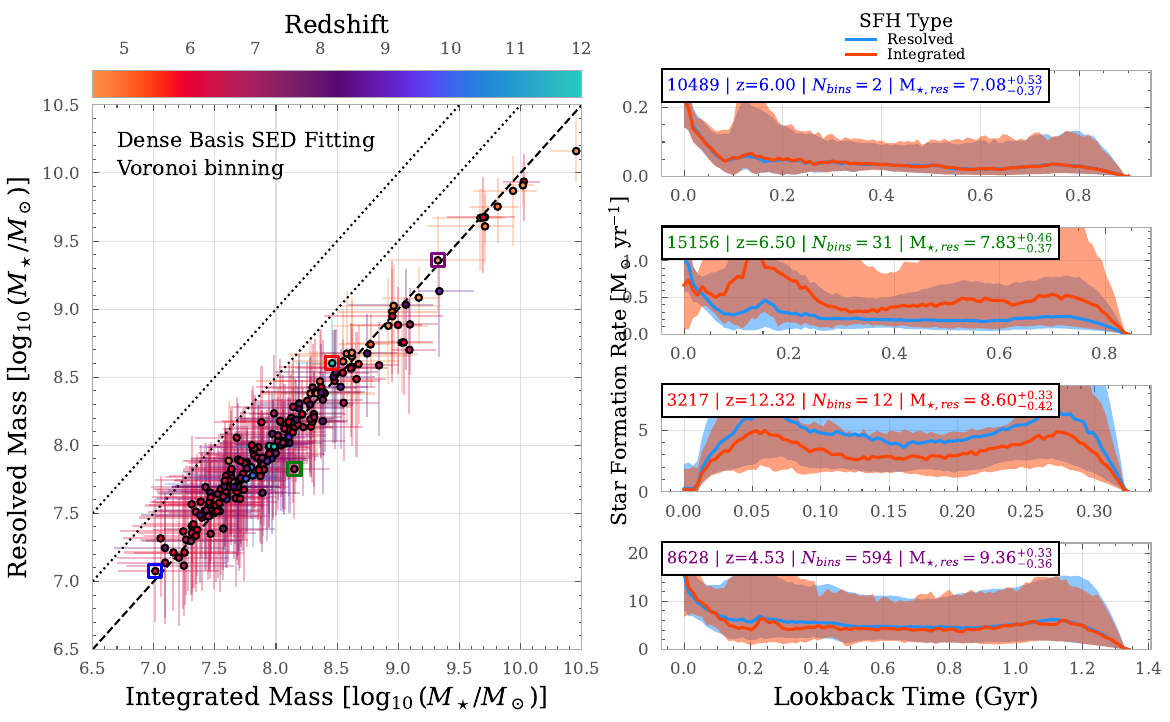}
    \caption{Comparison of resolved and integrated stellar masses from \db{} fitting, with four example inset star formation history plots for a representative spread of galaxies. The resolved fitting was performed on the `Voronoi' binned galaxy regions, and the resultant stellar masses for all individual regions were summed to calculate a total `resolved' stellar mass, with uncertainties calculated from the posterior mass distributions for each fit. for The point colour corresponds to the photometric redshift of the galaxy, and the coloured boxes highlight the position of the galaxies whose inferred star formation histories are shown. We can see that in the vast majority of cases the inferred stellar masses and estimated SFHs are consistent for both the integrated and resolved fits, which would suggest that outshining has little impact on the \db{} derived stellar masses.
    However, the SFH prior appears to favor long, smooth SFHs in almost all cases, even for low-mass galaxies, where we might expect more stochastic SFHs, which suggest the inferred SFHs may be prior driven in some cases and could be overestimating stellar masses for galaxies which are in face dominated by young stellar populations.}
    \label{fig:db_results}
    
\end{figure*}

\autoref{fig:db_results} shows the relation between the resolved and integrated stellar mass estimates derived from the \db{} fitting. We convert the total mass reported by \db{} to a surviving stellar mass comparable to those we report from \bagpipes{} using the assumptions of \cite{Madau2014} for a \cite{2003PASP..115..763C} IMF. It is evident from the low amount of scatter in this figure that \db{} consistently recovers the same inferred stellar mass for the vast majority of galaxies whether they are fit in an integrated or resolved manner. The uncertainties in stellar mass are derived from 16th and 84th percentiles of the posterior distributions, but appear systematically overestimated. 

The same figure highlights the recovered SFHs for 5 galaxies across a range of masses, showing in most cases a similar SFH for each galaxy is recovered. We highlight in particular ID \textit{8628}, which is shown on the bottom axis, where essentially identical SFH is recovered whether it is broken into 600 individual components or fit as a single integrated galaxy. 

When we compare the (total) resolved stellar masses estimated with \bagpipes{} and \db{} they are generally consistent for galaxies above 10$^9$ M$_\odot$ within the systematic uncertainties imposed by differing SPS models and slightly different IMFs. At lower masses \db{} produces systematically higher stellar masses than our resolved \bagpipes{} estimates, by around $\sim 0.5$dex, as the majority of galaxies are found to have significantly longer SFHs than inferred by our resolved SED fitting with \bagpipes{}. This could cause a potential overestimation of stellar mass for some low-mass galaxies, as this is a larger offset than the correction for outshining for most of these galaxies. 

It is also apparent from the highlighted SFHs, which are generally representative of the SFHs recovered for these galaxies, that \db{} tends to infer long periods of ongoing star formation history even in low-mass galaxies, where simulations and other observations may suggest more stochastic SFH histories and a shorter overall timescale for star formation \citep{faucher2018modelLooser2023,asada2023bursty,Endsley2023,2024ApJ...961...73N}. \cite{Iyer2019NonparametricFormation} note that the modelling of older stellar populations is typically prior rather than likelihood dominated. There may be cases where the vast majority of stellar mass in a galaxy has indeed been formed in a recent burst, and the inference of a non-existent older stellar population results in an overestimation of the stellar mass. We also observe differences in the inferred SFHs with the \db{} SFH implemented in \bagpipes{} to those fitted with \db{} directly, with the \bagpipes{} results typically showing shorter overall timescales and more variation from the assumed prior, which may be due to differences in implementation, posterior sampling or other model components. 

Whilst this consistency is encouraging we caution that as we do not know the true SFHs for these galaxies, we cannot say from this fitting alone whether \db{} is accurately recovering the star formation histories of these galaxies.  

We have not attempted to test other commonly used SED fitting tools such as Prospector \citep{Johnson2021} or BEAGLE \cite{chevallard2016modelling}, in part due the computational expense of fitting such a large number of galaxies with these tools, which take considerably longer per fit than \bagpipes{} or \db{}. 
Speed of inference is highly important when considering resolved pixel-by-pixel or binned SED fitting due to the number of fits, and performant SED fitting tools designed for resolved analysis (such as fitting within \pixedfit{}, \citealt{2022ApJ...935...98A}), with features such as informative priors based on neighbouring bins/pixels, or other novel ways of combining integrated and resolved analysis on different scales, may be helpful in enabling a more widespread and routine use of resolved SED fitting. 

\subsection{Inferred Effect of Outshining on GSMF and Stellar Mass Density}

In \cite{2025ApJ...978...89H} we used the EPOCHS galaxy sample of $\geq$1100 galaxies  \citep{Conselice_2025} to infer the galaxy stellar mass function (GSMF) at $z \geq 7$. The GSMF describes the overall distribution of galaxies as a function of stellar mass, and is typically parametrized by a Schechter function \citep{Schechter1976}, which is a power-law with an exponential cut-off. We explored the systematic effects caused by SED fitting on the GSMF and the resultant stellar mass density (SMD), which is the integral of the GSMF. 

As we have shown that outshining can cause a systematic offset in stellar mass for galaxies with $\leq 10^9 \; {\rm M_{\odot}}$, here we test whether the mass-offsets we observe will have a significant impact on the stellar mass function we derived in \cite{2025ApJ...978...89H}. We do not attempt to directly derive a new GSMF from the galaxies in this sample due to the relatively small cosmic volume probed, but rather we assume a systematic binned mass offset which we apply to our fiducial GSMF from \cite{2025ApJ...978...89H}. 

Our fiducial GSMF assumed a lognormal SFH, so we use this integrated SFH model and our Voronoi binning derived resolved fitting results when calculating the outshining correction to apply to our galaxy masses. We limit ourselves to the $6.5 \leq z <  7.5$ regime, as it is the lowest redshift bin of our GSMF, and at higher redshifts we do not have enough galaxies in this sample to derive a correction factor. Our sample contains 43 galaxies in this redshift range, from which we derive a binned correction factor as a function of integrated stellar mass for the low mass bins of our $z=7$ GSMF, which are $7 \leq \log_{10}\rm M_\star/M_\odot < 7.75$ and $7.75 \leq \log_{10}\rm M_\star/M_\odot < 8.5$. We do not apply any correction to galaxies in higher stellar mass bins, as we find consistent stellar masses for these galaxies. In the two bins we apply correction factors to we calculate median stellar mass offsets of 0.3, and 0.23 dex, with standard deviations of 0.25 and 0.3 dex respectively. We then choose to parametrize our correction factor as a log-normal distribution, in order to account for the rare cases where resolved and integrated mass estimates differ by 1 dex or more, which would not otherwise be captured with a simple linear correction. We convert our estimated median $\mu_{\rm M}$ and standard deviation ($\sigma_{\rm M}$ to the equivalent $\mu$ and $\sigma$ of a log-normal distribution such that $\mu = \ln(\mu_{\rm M}^2/\sqrt{\mu_{\rm M}^2 + \sigma_{\rm M}^2})$ and $\sigma = \ln(1+\sigma_{\rm M}^2/\mu_{\rm M}^2)$, finding $\sigma =0.76$, $\mu = -1.39$ and $\sigma =0.58$, $\mu = -2.33$ for the two bins respectively.

As our GSMF is bootstrapped such that we make many draws from the mass posterior for each galaxy, we apply our derived correction by drawing a correction factor from the log-normal distribution for each draw from the mass posterior and add them together to produce a corrected mass draw. We limit the maximum correction factor to the maximum offset seen in each bin (1.5 and 0.75 dex) in order to avoid drawing unphysically large correction factors from the tails of the distribution. We emphasize that is is a very approximate method and we are not accounting for the correlation of the stellar mass offset with properties such as burstiness or sSFR and are assuming it applies equally to all galaxies in a given mass and redshift bin. The model we assume here may slightly overestimate the effect of outshining and can be taken as a pessimistic case, as it does not account for the fraction of galaxies which scatter to lower resolved stellar masses (typically by less than 0.1/0.2 dex) than the integrated measurement. We are also implicitly assuming that the observed mass offset between our integrated and resolved stellar mass estimates for this sample holds for our integrated stellar masses measured from photometry for the full EPOCHS sample with different filtersets and depths, which is not necessarily always true, as we typically do not have as many medium-band observations available for the majority of the galaxy sample from \cite{2025ApJ...978...89H}. 

\begin{figure}
    \centering
    \includegraphics[width=\columnwidth]{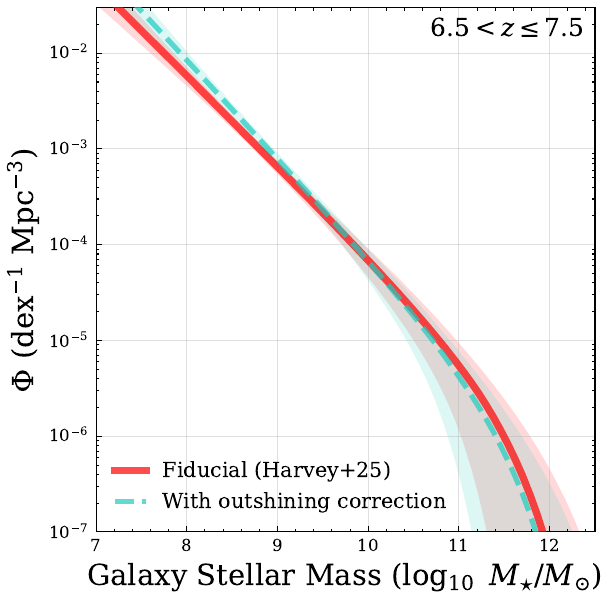}
    \caption{Inferred GSMF with our simplistic outshining correction at $z=7.0$, with a comparison to the fiducial model of \protect\cite{2025ApJ...978...89H}. The low-mass slope $\alpha$ shows a slight steepening from -1.94$^{+0.10}_{-0.10}$ to $-2.02^{+0.10}_{-0.10}$, but is still consistent with our fiducial fit within $1\sigma$. We find $\log_{10} \phi^{\star} = -6.08^{+1.06}_{-0.74}$ and $\log_{10}\rm M^\star = 11.55^{+0.65}_{-0.99}$ for the normalization and knee of this corrected GSMF. }
    \label{fig:gsmf}
\end{figure}

The resultant GSMF is consistent with our previous fiducial result within the estimated uncertainties. We observe a 40\% increase in volume density in our $7.75 \leq \log_{10}\rm M_\star/M_\odot < 8.5$ bin, but this is consistent within the uncertainties and this low mass bin is dependent on our completeness correction. We allow galaxies with stellar masses below our low-mass bin limit to scatter into a higher bin with the inferred outshining correction, which results in a net average increase of galaxies contributing to this bin from 227 to 284. 

We sample the posterior of a Schechter function fit to this GSMF using a Markov-Chain Monte Carlo sampler {\tt emcee}, which we plot with a comparison to our fiducial GSMF in \autoref{fig:gsmf}. The best-fit low-mass slope $\alpha$ shows a slight steepening from -1.94$^{+0.10}_{-0.10}$ for our fiducial model to $-2.02^{+0.10}_{-0.10}$ for the corrected model, but is still consistent with our fiducial fit within one $\sigma$.

 When we integrate our best fit model between 10$^8$ M$_\odot$ and 10$^{13}$ M$_\odot$ in order to calculate the cumulative stellar mass density, we find a value which is slightly higher than our previous measurement, but  consistent within the uncertainties, of $ \log_{10} \rm M_\odot/Mpc^3 = 6.38^{+0.12}_{-0.15} $. 

Most galaxies do not move mass bins, given the bin width of 0.75 dex used, and even those which scatter by more than this amount are only scattering to a maximum mass below $10^9$ M$_\odot$, where the impact of individual galaxies on the GSMF is small.
As we see a systematic increase in the effects of outshining with redshift, it may be possible that the effects on the GSMF are more significant at $z>8$, but we do not have a large enough sample to derive a reasonable correction at higher redshifts. The maximum age of an hidden `old' stellar populations is also decreasing with redshift, and there are considerable other systematic effects on SED derived stellar masses which make high-$z$ mass functions very uncertain \citep{2021MNRAS.501.1568F,2025ApJ...978...89H, weibel2024gsmf}. Hidden stellar mass in red galaxies with weak UV emission which are missed in a UV selected sample are likely have a larger systematic effect on the GSMF \citep{gottumukkala2023unveiling, 2024ApJ...968...34W, barro2024extremely}.

\subsection{Limitations}

In this section we will briefly discuss potential caveats of our results and analysis. Whilst we have attempted to mitigate potential problems and ensure our results are as accurate as possible, it is important to be honest about the limitations of the available data and our analysis. 

Firstly we must consider the impact of our binning procedures on our resolved SED fitting results. In the cases where we bin pixels when SED fitting, we are potentially collating regions of young and old stars, so individual bins can still be subject to the effects of outshining. Even in the case of pixel-by-pixel fitting, we are still collating stellar populations on the order of hundreds of parsecs, which may contain a mix of both old ($\geq 100$ Myr) and young ($\leq$ 10 Myr) stellar populations. The effect of the PSF also means that the light from these stellar populations is also spread over many pixels, so a small region of intense star formation with strong line emission will affect the measured photometry for surrounding regions. That is to say that even the pixel-by-pixel case is likely a lower-limit on the stellar mass due to this unresolved outshining and may partially explain the larger mass offsets seen by \cite{gimenez-artega2022, giménezarteaga2024outshining} given the spatial magnification provided by strong lensing.

One other potential issue is the lack of rest NIR observations for these galaxies. \cite{song2023solution} shows at $z<3$ if photometric data redder than $1\mu$m rest-frame is not available, stellar masses can be systematically overestimated by up to 0.2 dex when SED fitting, which could explain some of the systematic bias we observe. They do not consider the improvement offered by medium-band photometry in their study however, which may partially mitigate this effect by allowing better constraints on emission lines. Obtaining rest-frame 1-1.6$\mu$m data at these wavelengths is extremely challenging due to the lower sensitivity of MIRI. For a $z = 10$ galaxy, we can see only the rest-frame UV and a fraction of the optical (to around 4500\AA{}), which is well below the wavelengths at which older stars will dominate the continuum. We can see that some of our largest mass discrepancies between integrated and resolved models occur in galaxies at $z=10$ or above, and this may be because of increased uncertainty in the fitted SFH rather than an inherent change in the underlying star formation rate and stellar populations. However we also note that both our resolved and integrated measurements cover the same wavelength range and this outshining comparison does not necessarily need to recover the true mass of each galaxy, only the relative mass difference.

One potential limitation of a pixel-by-pixel analysis, which is somewhat mitigated by a binned analysis such as our Voronoi binning, is the potential to exclude older stellar populations on the outskirts of galaxies due to low surface brightness, leading to individual pixels not meeting imposed SNR criteria. Binning these pixels together can allow us to recover more of this signal, if the regions that are binned are reasonably contiguous. 

For bins/pixels that are close to our SNR limits, there is a potential risk of prior-driven, rather than posterior constrained SFHs. We have not tested varying our prior assumptions for our resolved SFH models. We do not have a ground truth to compare to and are making the explicit assumption in our analysis that our resolved results are more accurate than the integrated analysis. There may however be cases where the individual fits are not well-constrained and we are recovering our prior, or indeed cases where our resolved fits are not accurately recovering the SFH of the galaxy due to some systematic or modelling effect we have not considered e.g. such as more complex dust-law or higher than estimated RMS noise.






%


\section{Conclusions}
\label{sec:conclusions}

In this work we perform spatially resolved SED fitting for a sample of 222 galaxies at $z>4.5$ in the Jades Origin Field. We test a variety of spatial binning methods, and build feature maps for each galaxy for properties including the stellar mass, star formation rate, and stellar age. We compare the resolved and unresolved mass estimates in order to search for evidence of outshining, where young stellar populations outshine older stellar populations. 

In detail, the key findings of this work are as follows.

\begin{enumerate}
    \item We find evidence for systematic underestimation of stellar mass in galaxies with M$_\star <10^9$ M$_\odot$, due to the outshining of older stellar populations by recent star formation when performing integrated SED fitting. The systematic mass offset observed differs slightly with the chosen integrated SFH models, but appears to increase with redshift for all tested SFH models, which may be due to more bursty star formation histories or be driven primarily by a selection effect. 
    \item We observe a minority of extreme cases where the mass discrepancy between resolved and integrated stellar masses due to outshining exceeds 1 - 1.5 dex, which occurs at high sSFRs during bursts of recent SFH. The median offset for galaxies below 10$^9$ M$_\odot$ is typically only 0.1 to 0.3 dex dependent on the binning and SFH model used, which is in line with other systematic SED fitting effects such as the assumed dust law or SPS model. 
    \item We see a weak correlation of mass offset with inferred H$\alpha$ EW, which we estimate from a photometric excess. Whilst the largest mass offsets observed do typically have inferred EWs $\geq$1000\AA{}, some of the highest EWs observed show little offset between resolved and integrated masses, with consistent SFH estimates in both fitted cases. 
    \item At these redshifts we find that burstiness (SFR$_{\rm 10 \ Myr}$/SFR$_{\rm{100 \ Myr}}$) shows a stronger correlation with the amount of outshining rather than sSFR alone, as a large fraction of galaxies with high sSFRs ($>10^{-8}~ \rm yr^{-1}$) show consistent masses for both the resolved and integrated SED fitting results. 
    \item The integrated SFH model which produces the least discrepant stellar mass estimate when compared to our resolved SED fitting depends on the choice of binning. In the cases with the largest number of bins per galaxy we find SFH models which allow older stellar populations, such as the `continuity' model of \cite{leja2019measure} or a double power-law SFH provide the least-discrepant integrated stellar mass estimate in most cases. However we observe a variation between galaxies such that all tested integrated SFH models produce the least-discrepant stellar mass estimate in at least a few cases. 
    \item We compared our initial case of a simple constant SFH for each resolved bin/pixel to the more flexible non-parametric `continuity bursty' model of \cite{tachella2022} in order to test whether the stellar mass of individual components varies when allowed a more complex SFH. We find overall a slight systematic decrease in our resolved stellar masses ($\approx$ 0.1 dex) for the more complex model, but our overall conclusions about outshining do not change when we substitute our resolved SFH model. 
    \item We test an alternative SED-fitting code, \db{}, to see if it can more accurately reproduce star formation histories between the integrated and resolved cases. We find that it produces more consistent stellar masses in both cases, but the resolved estimates from \db{} are typically systematically offset $\geq 0.5$ dex higher than the resolved stellar masses we infer with \bagpipes{}, particularly at $\leq 10^9$ M$_\odot$. The vast majority of these low-mass galaxies are fitted with long SFHs, which may be driven by the SFH prior rather than constrained by the fit itself, and may be systematically overestimated. 
    \item We find that NIRCam medium-band observations of comparable depth to the widebands can decrease the impact of outshining by providing better constraints on the stellar continuum and emission lines for fitting integrated photometry when compared to SED fitting using wide-band only photometry, and they also reduce overall scatter in stellar mass by $\sim$0.5~dex.  
    \item We derived a correction factor to account for the observed outshining and applied it to our fiducial $z=7$ GSMF in \cite{2025ApJ...978...89H}, finding a 40\% increase in $\phi$ for our 10$^{7.75 - 8.5}\rm M_\odot$ and a slight steepening in the best-fit Schechter function low-mass slope $\alpha$ from -1.94$^{+0.10}_{-0.10}$ to $-2.02^{+0.10}_{-0.10}$, but overall the fitted GSMF is consistent with our fiducial model within $1\sigma$. We suggest that whilst outshining can have a large impact on individual galaxies, the overall effect on the galaxy mass distribution and the total stellar mass density is small. 

Galaxies are complex structures, and spatially-resolved SED fitting is an important technique to fully understand variations in stellar populations, dust content and nebular emission. The resolution of JWST makes this possible at $z>6$ for the first time. We encourage future studies to consider the spatial variation of high-redshift galaxies where possible in order to increase the accuracy of derived parameters. 

\end{enumerate}

\section*{Acknowledgements}

TH, CC, NA, DA, QL, LW, VR and CMG acknowledge support from the ERC Advanced Investigator Grant EPOCHS (788113), as well as two studentships from the STFC. CCL acknowledges support from a Dennis Sciama fellowship funded by the University of Portsmouth for the Institute of Cosmology and Gravitation. RKC is grateful for support from the Leverhulme Trust via the Leverhulme Early Career Fellowship. APV acknowledge support from the Sussex Astronomy Centre STFC Consolidated Grant (ST/X001040/1). JT acknowledges support from the Simons Foundation and JWST program 3215. Support for program 3215 was provided by NASA through a grant from the Space Telescope Science Institute, which is operated by the Association of Universities for Research in Astronomy, Inc., under NASA contract NAS 5–03127.

We thank Geferson Lucatelli for providing the computational resources necessary to perform this analysis. 
We would like to thank John Weaver, Sam Cutler and Katherine Whitaker for publicly distributing the {\tt aperpy} code \citep{2024ApJS..270....7W}, upon which we base our PSF modelling. We would also like to thank Abdurro’uf and Dan Coe for publicly releasing \pixedfit{}, as well as Adam Carnall for \bagpipes{} and Kartheik Iyer for \db{}.

This work is based on observations made with the NASA/ESA \textit{Hubble Space Telescope} (HST) and NASA/ESA/CSA \textit{James Webb Space Telescope} (JWST) obtained from the \texttt{Mikulski Archive for Space Telescopes} (\texttt{MAST}) at the \textit{Space Telescope Science Institute} (STScI), which is operated by the Association of Universities for Research in Astronomy, Inc., under NASA contract NAS 5-03127 for JWST, and NAS 5–26555 for HST. The authors thank all involved with the construction and operation of JWST, without whom this work would not be possible. The authors also thank the team who designed and proposed for the JADES (PID 1280,1287) and JADES Origins Field (PID 3215) observing programmes, whose data was used in this study. Elements of this work also used observational products from CANUCS, GLASS, CEERS, FRESCO, JEMS, NGDEEP, PEARLS and MegaScience JWST surveys. We thank all of those involved in these collaborations for their efforts in designing and carrying out these observations. 


This paper has made use of Python libraries including {\tt astroquery \citep{ginsburg2019astroquery}, astropy \citep{astropy:2013, astropy:2018, Astropy2022}, bagpipes \citep{Carnall2018_Bagpipes, carnall2019vandels}, bokeh \citep{jolly2018hands}, cmasher \citep{van2020cmasher}, corner \citep{corner}, dense\_basis \citep{Iyer2019NonparametricFormation}, emcee \citep{emcee}, fsps \citep{conroy2010fsps}, galfind \citep{2024arXiv240410751A}, h5py \citep{collette2023h5py}, hvplot \citep{yang2022holoviz}, jwst \citep{Bushouse2022}, matplotlib \citep{Hunter2007}, numpy \citep{harris2020array}, numpyro \citep{phan2019composable}, opencv \citep{bradski2000opencv}, pandas \citep{reback2020pandas}, panel, photutils \citep{larry_bradley_2022_6825092}, piXedfit \citep{2021ApJS..254...15A,2022ApJ...926...81A,2022ApJ...935...98A}, pymultinest \citep{buchner2016pymultinest}, pypher \citep{boucaud2016convolution}, pysersic \citep{pasha2023pysersic}, scikit-image \citep{van2014scikit}, scikit-learn \citep{pedregosa2011scikit}, scipy \citep{2020SciPy-NMeth}, sep \citep{barbary2016sep}, synthesizer \citep{FLARES-II,10.1093/mnras/staa649}, webbpsf \citep{Perrin2012,Perrin2014} and xarray \citep{hoyer2017xarray}}.

\section*{Data Availability}

 The JWST data used in this analysis is available to download from MAST at \url{https://archive.stsci.edu/}. 
The code used for the analysis in this paper is publicly available on GitHub at \url{https://github.com/tHarvey303/EXPANSE}. The code to generate the figures in this manuscript is available in a Jupyter notebook at \url{https://github.com/tHarvey303/EXPANSE/blob/master/scripts/outshining_paper_figures.ipynb}. Catalogues of results and the data files will also be provided upon reasonable request.



\bibliographystyle{mnras}
\bibliography{main} 




\appendix


\section{PSF Modelling}

\label{sec:psf}

Given that our spatially-resolved analysis operates close to the scale of the NIRCam Point Spread Function (PSF), accurate modelling of the PSF is critical to ensure reliable results. PSF homogenization is carried out to ensure reliable photometry in all photometric bands, given the broadening of the PSF with increasing wavelength. We homogenize all our imaging to F444W, which has the broadest PSF of the available filters. 

The NIRCam PSF can vary significantly with time, due to variation in the alignment of the 18 individual mirror segments. Wavefront sensing calibrations, which are measured on orbit every few days, can be used to model the PSF of NIRCam close to the epoch of a given observation using the \webbpsf{} tool \citep{Perrin2012, Perrin2014}.

PSF modelling can also be carried out by stacking isolated point sources, typically stars, in the image. Numerous studies have found that the NIRCam PSF models produced from stacking stars are slightly broader than the simulated PSF models produced by \webbpsf{}, even when accounting for the effect of drizzling during the reduction process \citep{2024ApJ...963....9M, 2024ApJS..270....7W}.

\cite{2024ApJ...963....9M} found that increasing the {\tt jitter\_sigma} parameter in \webbpsf{}, which approximates telescope dither, from 0.007 to 0.022 (0.034) arcsec for the SW (LW) detectors, better reproduced the PSF derived from stacked stars. 

We derive empirical PSFs from stacking stars in the imaging, following the methodology of \cite{2024ApJS..270....7W}, using a modified version of a routine from their software package \aperpy\footnote{\url{https://github.com/astrowhit/aperpy/}}. 

This methodology identifies stars as point sources which occupy a defined region within a size-magnitude plane, where the F$(<0\farcs16)/$F$(< 0\farcs32)$ ratio is used as a proxy for size, and magnitudes measured within a $0\farcs32$ diameter apertures. Stars are identified by fitting a slope to this plane, selecting those with apparent magnitudes $18 <  $m$_{\textrm{AB}} < 25$, sigma clipped at $<3.5\sigma$. This typically results in 10 - 15 stars per NIRCam image. We make 4$\arcsec$ cutouts for each star, which are centered, normalized and stacked. Cutout with centres of mass more than 3.5 pixels (0\farcs1) from the initial identified peak are excluded for potential contamination by neighbours. The stacking procedure uses sigma-clipping on a pixel-by-pixel basis, excluding pixels more than 2.8$\sigma$ from the mean of that pixel. The PSFs are then normalised to the tabulated enclosed energies measured during NIRCam calibration in a 4\arcsec diameter\footnote{\url{https://jwst-docs.stsci.edu/jwst-near-infrared-camera/nircam-performance/nircam-point-spread-functions}} \footnote{\url{https://www.stsci.edu/hst/instrumentation/acs/data-analysis/aperture-corrections}}. 

For the ACS/WFC we use the full GOODS-South mosaic to identifying stars in order to construct more reliable PSF models. We do not consider the variation in PSF across the ACS/WFC fields due to the inhomogeneous nature of the mosaics, which combine exposures at many position angles over a baseline of 12 years. This is consistent with the approach taken by \cite{Whitaker2019} for the public GOODS-South catalogue.

\begin{figure*}
    \centering
    \includegraphics[width=\textwidth]{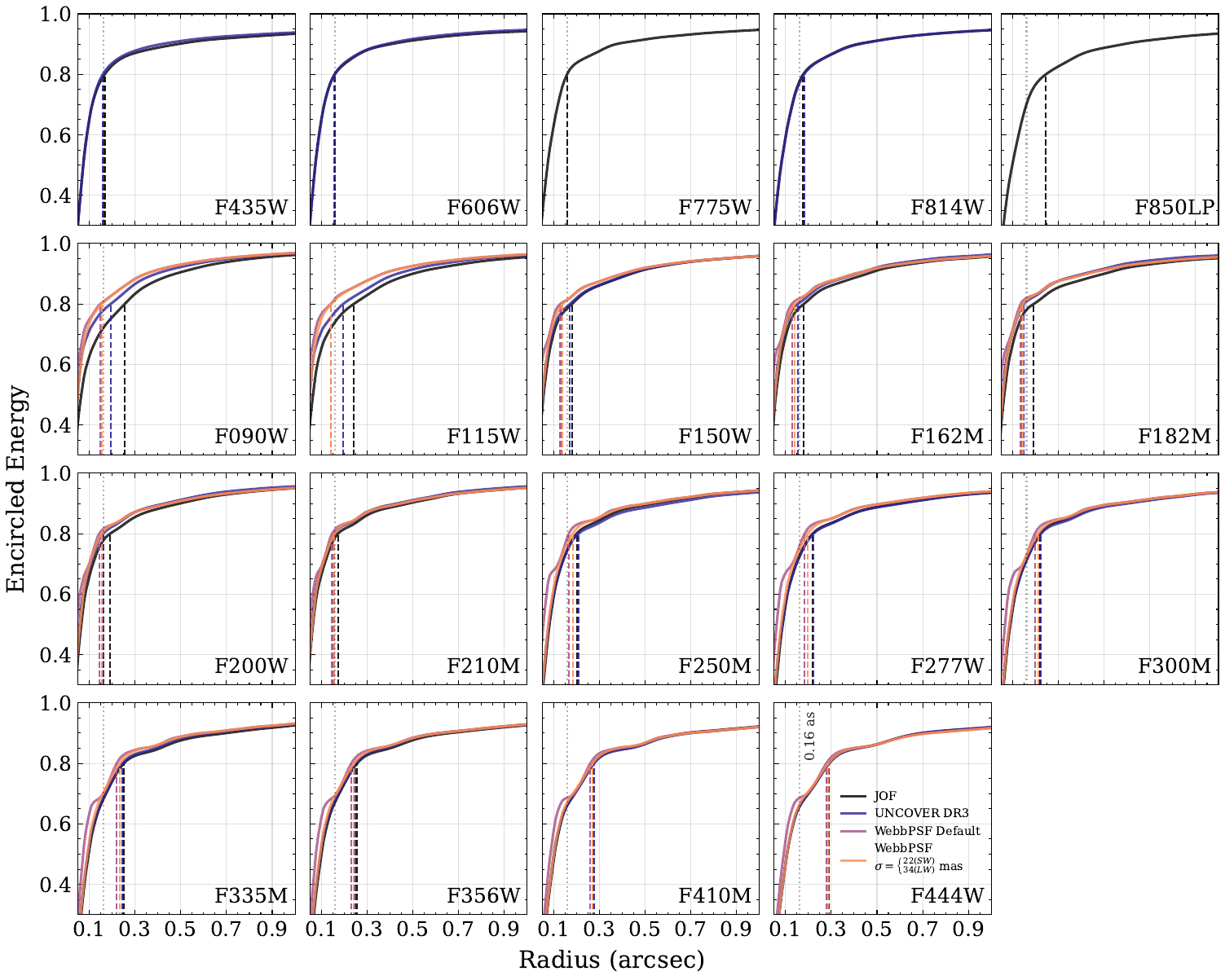}
    \caption{Comparison of encircled energy between PSF models derived from \webbpsf{} and empirically from stacking stars using the methodology of \protect\cite{Whitaker2019, 2024ApJS..270....7W}. We show both the default \webbpsf{} model as well as the proposed variation of  \protect\cite{2024ApJ...963....9M}, with the {\tt jitter\_sigma} parameter of 22 (34) mas for the long and short wavelength bands respectively. We also compare to the publicly released convolution kernels from  \protect\cite{2024ApJS..270....7W} and  \protect\cite{suess2024medium} from the UNCOVER field. Vertical dashed lines show the 80\% enclosed flux radius for each model.}
    \label{fig:psf_ee}
\end{figure*}

\autoref{fig:psf_ee} shows the encircled energy as a function of radius for all filters, with a comparison to \webbpsf{} models and publicly released PSF models for other fields. We find slightly broader PSFs than the default WebbPSF models, but are generally consistent with the broadened WebbPSF models of \cite{2024ApJ...963....9M} and the publicly released models by the UNCOVER/MEGASCIENCE teams \citep{2024ApJ...963....9M,suess2024medium}, with the exception of F090W and F115W, where our PSF models are slightly broader. All stacked cutouts for these filters were checked by eye and found to be visually consistent with isolated point sources. 

We use {\tt PYPHER}, with a regularisation parameter of 1e-4, to produce convolution kernels between our longest wavelength filter, F444W, and all other filters \citep{boucaud2016convolution}. Convolution kernels are produced at a $3\times$ oversampling and then rescaled to our standard 0$\farcs03$ pixel scale. 

\begin{figure}
    \centering
    \includegraphics[width=\columnwidth]{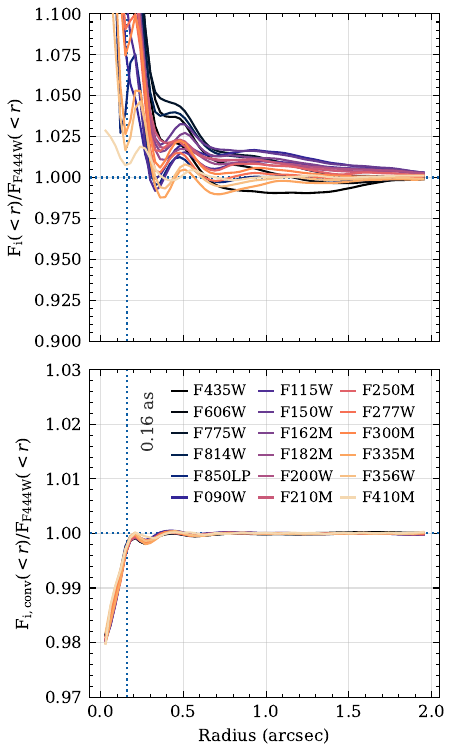}
    \caption{PSF growth curves relative to F444W before and after convolution with our derived kernels using {\tt pypher}. The consistency of our convolution kernels can be seen, although we note the 2\% deviation in flux at radii below 0.16 arcsec.}
    \label{fig:psf_rel_cog}
\end{figure}

The PSF growth curves for all filters before and after convolution are shown in \autoref{fig:psf_rel_cog}, with very small deviations ($<1\%$) within our initial 0$\farcs$32 apertures. At the smallest scales we see a small systematic decrease which is likely to have a small residual effect on our pixel-binning analysis in later sections. We discuss the impact of using the \webbpsf{} derived model instead of a empirical model in Appendix \autoref{sec:webbpsf}.

{}

\section{EXPANSE Overview}
\label{sec:expanse}

\begin{figure*}
    \begin{subfigure}[b]{0.80\textwidth}
    \includegraphics[width=\textwidth]{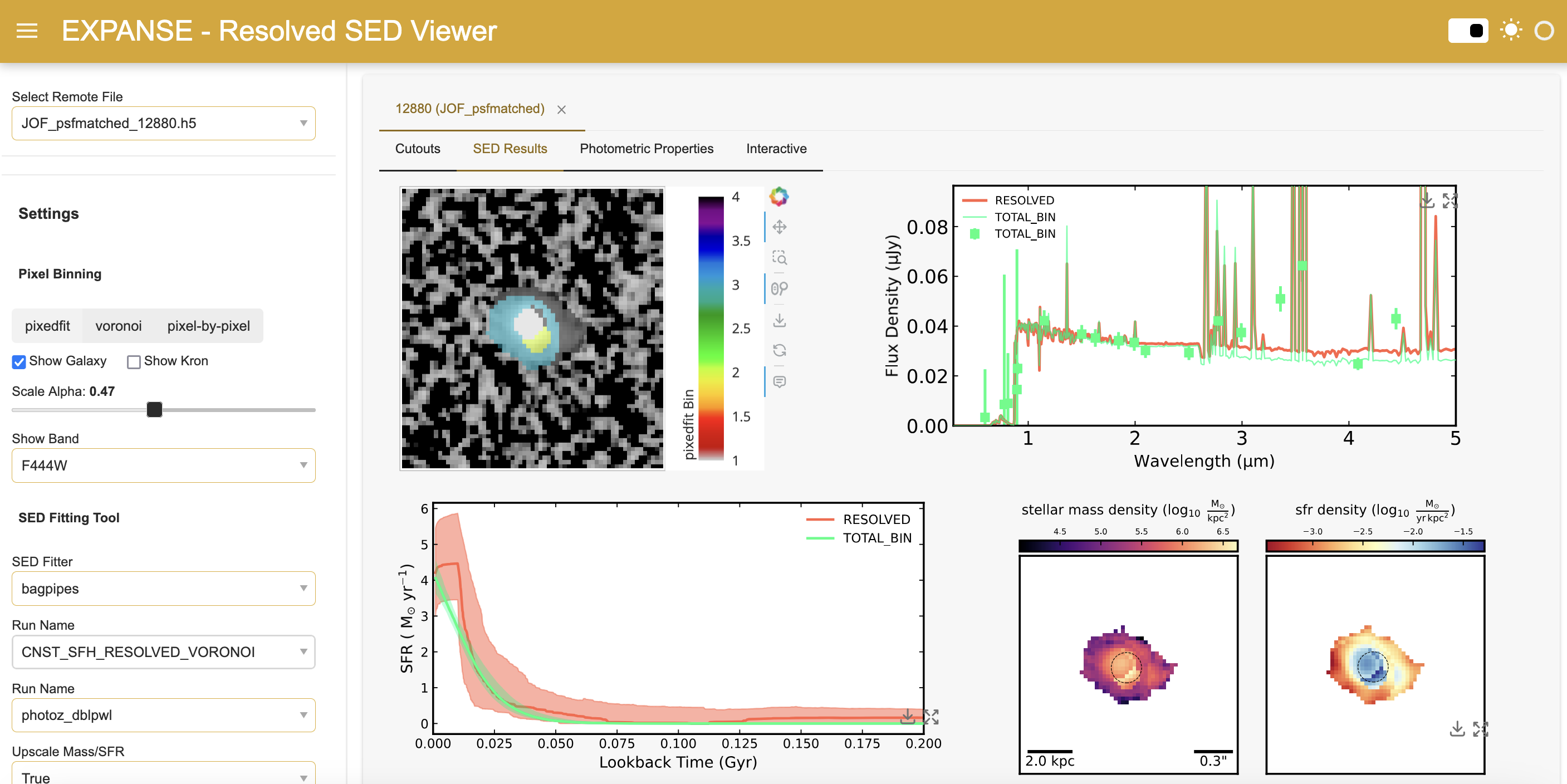}
    \subcaption{Example of SED fitting results tab of \expanse{} Web UI, with sidebar controls to customize plots and show different results.}
    \end{subfigure}

    \begin{subfigure}[b]{0.80\textwidth}
        \includegraphics[width=\textwidth]{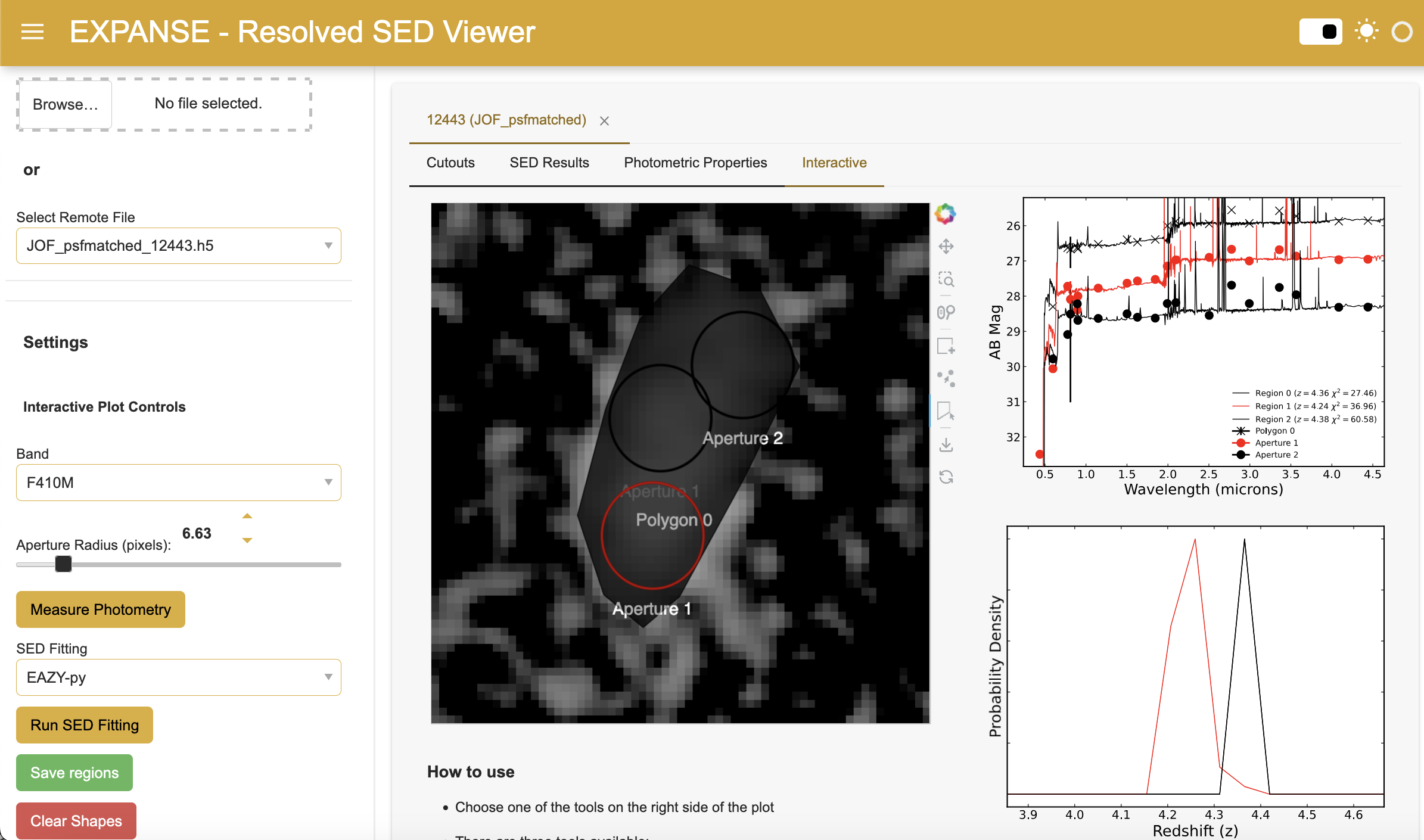}
        \subcaption{Example of interactive SED fitting tab, showing placeable apertures/polygon regions and \eazy{} SED fitting of the measured photometry.}
    \end{subfigure}
    
    \caption{Examples of \expanse{} GUI interface. }
    \label{fig:expanse_features}
\end{figure*}

\begin{figure*}
    \centering
    \includegraphics[width=\linewidth]{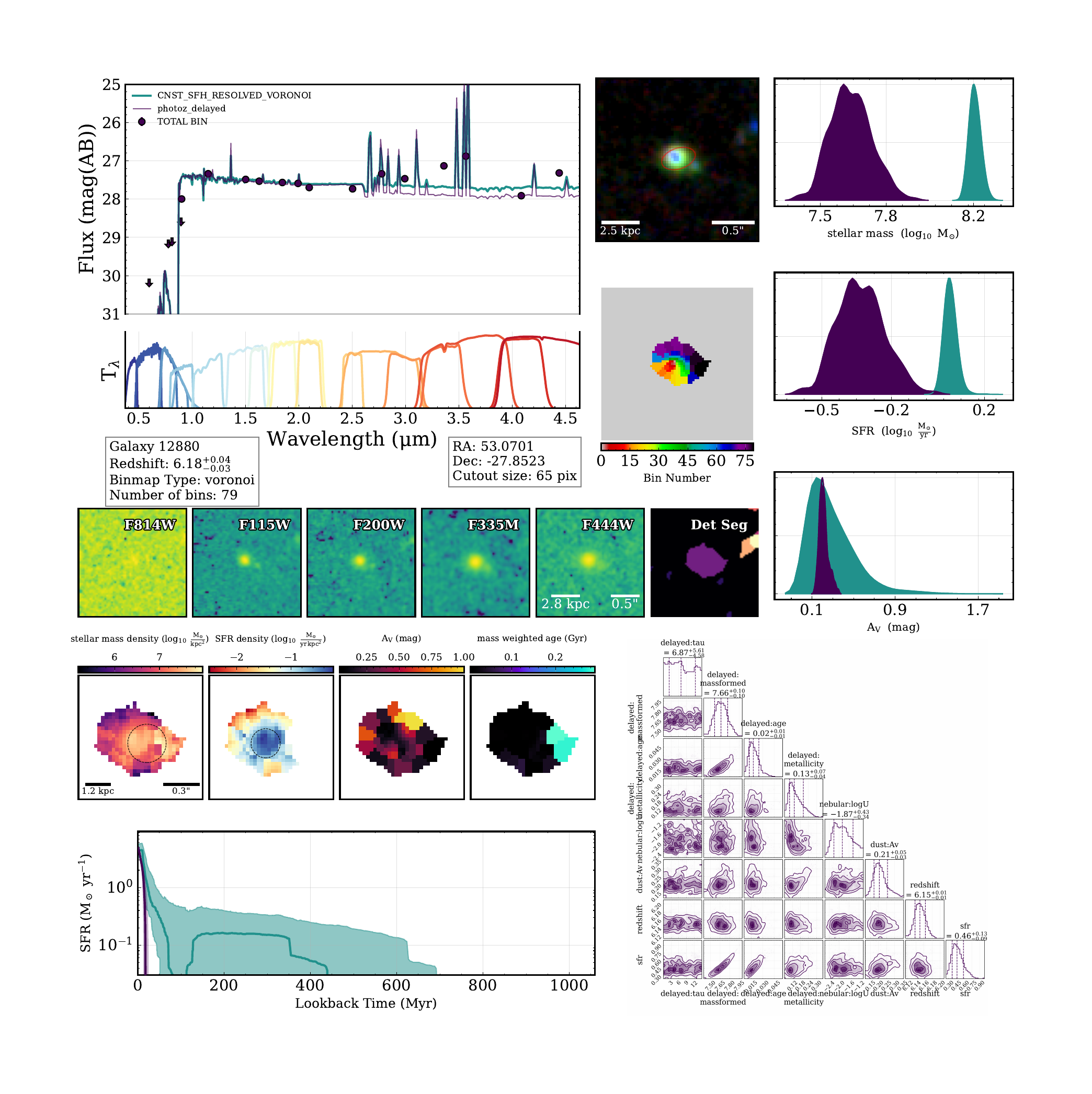}
    \caption{Example $z\sim6.2$ galaxy in the JOF showing information available from resolved SED fitting, including plots of the best-fitting SEDs, inferred SFHs, property maps and corner plots. In the SED, SFH and posterior plots the blue line shows the resolved fit, and the purple line shows the results of an integrated fit with a delayed-$\tau$ SFH. The right hand side shows the stellar mass and SFR posterior distributions for the combined resolved fits and example integrated fit. Apparent regional variations in properties such as dust attenuation and mass-weighted are apparent from the property maps and corner plots. The colour used for each bin as labelled in the upper SED panel is used consistently in the parameter posterior distribution, corner and SFH plots.}
    \label{fig:seds}
\end{figure*}

The resolved analysis we perform in this work is enabled through a software package we have developed called \textit{EXtended Pixel-resolved ANalysis of SEDs} (\expanse{}). We have publicly released \expanse{} alongside this paper, along with the code to produce the analysis and figures at \url{https://github.com/tHarvey303/EXPANSE}. \autoref{fig:expanse_features} shows some examples of the interactive GUI interface for \expanse{}. \autoref{fig:seds} shows an example of the resolved SED-fitting results which can be produced by \expanse{} for one of the galaxies studied in this paper. 

As this code has not been previously published, in this appendix we briefly highlight the key features of \expanse{}, including some features that are not used in this work. \expanse{} is a Python module which can be cloned from the GitHub repository and will install required dependencies during installation.

\begin{enumerate}
    \item Easy setup: \expanse{} ties into \galfind{}, so any galaxy from a catalogue processed with this tool can be automatically converted into the \expanse{} format. We also provide a helper class to allow initialization from just a set of image mosaics, a center coordinate and a cutout scale.
    \item Portability: All information for each galaxy is serialized into a compressed HDF5 file, and can be quickly loaded, transferred between computers, and backed up.
    \item Compatibility: \expanse{} is designed to support a wide array of external tools for SED fitting, binning, morphological fitting, galaxy detection and segmentation and other analysis.
    \begin{enumerate}
        \item Galaxy Detection: {\tt sep} and {\tt photutils} are currently supported, as is the ability to load abritary precomputed photometry and segmentation maps.
        \item SED-Fitting: \eazy{}, \db{} and \bagpipes{} are currently supported, with plans to add additional codes in the future.
        \item Binning: \pixedfit{}, {\tt vorbin}, as well as internal methods for pixel-by-pixel binning are implemented, and it is possible to extend this to arbitrary binning. 
        \item Morphological Fitting: \pysersic{}, {\tt pyautogalaxy} and {\tt pygalfitm} are supported for fitting one or more component Sersic or PSF models with complex priors. 
    \end{enumerate}
    \item: Interactability: \expanse{} supports conventional interaction via scripting or command line, as well
            as an interactive web-based GUI which allows you to load and interact with each galaxy, compare and overlay fit results for different bins or components, and place or draw arbitrary apertures for photo-$z$ estimation and SED fitting. 
    \item Testability: \expanse{} supports mock images generated from hydrodynamical simulations, and can generate mock images from particle distributions using {\tt synthesizer} and add realistic noise and the effects of the PSF. Ground truth measurements of SFH, spectra and parameters such as stellar mass, emission line fluxes and metallicity are stored for comparison of SED fitting estimates. 
    
\end{enumerate}

\section{Effect of Chosen PSF Model}
\label{sec:webbpsf}

As with any study of resolved galaxy properties, the chosen PSF model can have a significant effect. As discussed in \autoref{sec:psf}, we use empirical PSFs for homogenizing our images in this analysis. In this section we briefly report on the results of experimenting with our alternative PSF model, using \webbpsf{} with added jitter, which we compare in terms of encircled energy in \autoref{fig:psf_ee}. We do not repeat our full analysis, but instead randomly select a small subset of galaxies from our sample, and recreate the analysis using imaging which we PSF match with \webbpsf{} derived convolution kernels, which we derive from the simulated PSFs using {\tt pypher} in the same fashion as our empirical PSFs. Our WebbPSF models use the best fit {\tt jitter\_sigma} parameters from \cite{Morishita2022} as described in \autoref{sec:psf}. We calibrate specifically to the epoch of observation by obtaining the nearest wavefront sensing calibrations for each image, typically taken within 12-24 hours. The PSF can also vary slightly across the image plane, so for each galaxy we model the PSF specifically for the detector the galaxy was located in for that exposure. We neglect the HST filters from this analysis, as they are excluded from all binning criteria, and it is more difficult to obtain an accurate simulated PSF. 

As we mention in \autoref{sec:psf}, the largest discrepancy between our empirical and \webbpsf{} models occurs in the NIRCam SW bands, specifically F090W and F115W, where our empirical model is broader than the simulated PSF, even with the additional {\tt jitter\_sigma} added. When the PSF-matched cutouts are compared directly, this results in a significant reduction of flux in the most central 2-3 pixels, along with an excess in the surrounding region. For our \pixedfit{} binning this has a limited effect on our binning given the minimum bin diameter of 7 pixels we impose, but the `pixedfit nomin' and Voronoi binning method, typically resulting in a slightly lower number of bins (10-20\% less), as the more compact PSF means the outskirts of the galaxies are fainter and thus each bin is slightly larger to reach the SNR requirement. 

We rerun our fiducial resolved constant SFH SED fitting model for the `pixedfit', `pixedfit nomin' and Voronoi bins as described in \autoref{sec:sed_fitting} using the photometry extracted from our \webbpsf{} matched cutouts. We find a small systematic increase in total resolved stellar mass compared to the results with our empirical PSF models, with a median offset of 0.05 to 0.13 dex, with a larger offset seen with the Voronoi binning method, which has the largest number of bins. 
As this effect acts to increase the total resolved stellar mass it would increase the observed mass discrepancy due to outshining, suggesting our observed mass offsets may be lower limits when considering other PSF models. We caution that the effect on resolved stellar mass maps, or other derived properties which don't scale with total light are more complex and future studies should carefully consider the choice and accuracy of their PSF model when undertaking a resolved analysis.

\section{Resolved vs Integrated Measurements of SFR}
\label{sec:resolved_sfr}
\begin{figure*}
    \includegraphics[width=\textwidth]{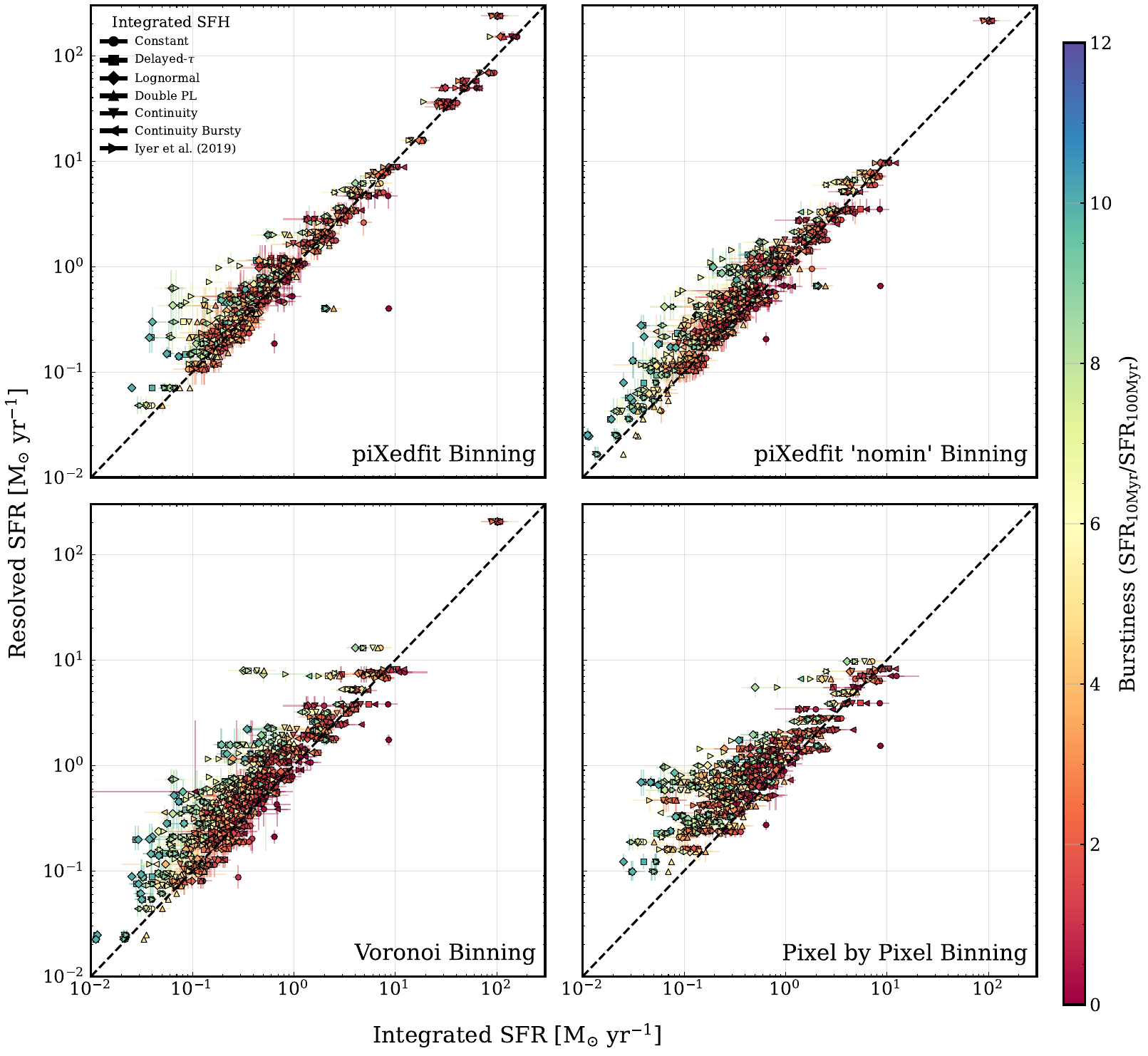}
    \caption{Resolved vs integrated SFR for 5 different \bagpipes{} integrated SFH models (shown with marker shape) and 4 different binning methodologies. Markers are coloured by the `burstiness' of the integrated SFH.}
    \label{fig:sfr_comp}
    
\end{figure*}

\autoref{fig:sfr_comp} shows a comparison of the integrated and resolved \bagpipes{} star formation rates for our four binning methods. The SFR here is averaged over a 100 Myr timescale. We see systematic underestimation of star formation rates with our integrated measurements, such that the integrated fits do not seem to be capturing the full star formation activity occurring, particularly at low SFRs. However there is a lower limit on the SFR constrainable from photometry alone, so the difference between estimates where the total SFR is $< 1$ M$_\odot$ yr$^{-1}$ is not necessarian significant, and we generally see reasonable agreement for higher SFRs which can be better constrained. 

\bsp	
\label{lastpage}
\end{document}